\documentclass[acmsmall, screen=true, nonacm=true]{acmart}

\usepackage{graphicx}
\usepackage{longtable}
\graphicspath{{figures/}} 
\usepackage{multirow}
\usepackage{subfigure}
\usepackage{colortbl}
\usepackage{url}
\usepackage{eurosym}
\usepackage{xcolor}
\usepackage{soul}
\usepackage{algorithm2e}

\AtBeginDocument{%
  \providecommand\BibTeX{{%
    \normalfont B\kern-0.5em{\scshape i\kern-0.25em b}\kern-0.8em\TeX}}}

\setcopyright{acmcopyright}
\copyrightyear{2018}
\acmYear{2018}
\acmDOI{XXXXXXX.XXXXXXX}

\acmJournal{JACM}
\acmVolume{37}
\acmNumber{4}
\acmArticle{111}
\acmMonth{8}

\begin{document}

\title{ElectAnon: A Blockchain-Based, Anonymous, Robust and Scalable Ranked-Choice Voting Protocol}

\author{Ceyhun Onur}
\email{ceyhun.onur@boun.edu.tr}
\orcid{0000-0002-0887-0786}
\author{Arda Yurdakul}
\email{yurdakul@boun.edu.tr}
\affiliation{%
  \institution{Bogazici University, Department of Computer Engineering}
  \city{Istanbul}
  \country{Turkey}
  \postcode{34342}
}


\begin{abstract}
Remote voting has become more critical in recent years, especially since the Covid-19 outbreak. Blockchain technology and its benefits such as decentralization, security, and transparency have given rise to proposals for blockchain-based voting systems. However, the traceability of blockchain transactions violates voter anonymity in existing proposals. Besides, transaction costs also need to be considered. Solutions that may cause repeated elections should be avoided for a low-cost scalable voting system.  In this work, we propose ElectAnon, a blockchain-based, self-tallying and ranked-choice voting protocol focusing on anonymity, robustness, and scalability. ElectAnon achieves anonymity by enabling voters to register with identity commitments and cast their votes via zero-knowledge proofs. Robustness is realized by removing the direct control of the authorities in the voting process by using timed-state machines. Each voter encodes the ballot into a single integer and blinds the vote off-chain while making the verification on-chain. This makes the protocol infinitely scalable in the number of voters. ElectAnon is also a solution for governance in Decentralized Autonomous Organizations (DAO): it includes a candidate proposal module and an algorithm-agnostic mechanism to plug-in different tallying methods easily. The \textit{Merkle forest} extension is proposed for conducting even more trustless elections. ElectAnon is implemented with smart contracts based on Ethereum Virtual Machine (EVM) and a zero-knowledge gadget, Semaphore. The implementation also includes two different sophisticated tallying methods, Borda Count and Tideman. Experimental results show that a 40-voter and 10-candidate election can be implemented with the gas consumption reduced up to 89\% compared to previous works. While other studies could not exceed a 25,000-voter setup, ElectAnon has been observed to run safely for 1,000,000 voters. The implementation can be found at \url{https://github.com/ceyonur/electanon}.
\end{abstract}



\keywords{elections, voting systems, i-voting, blockchains, distributed systems, smart contracts, zero-knowledge proofs}

\maketitle
\section{Introduction}
\label{chapter:introduction}
The recent COVID-19 pandemic has pushed existing systems and procedures to be implemented remotely. Legacy election systems, which oblige voters to present in a specific place to cast a ballot, have also been impacted by the pandemic. For example, almost half of the votes were cast through mail voting in the US 2020 presidential election \cite{desilver-mail-voting-nodate}. Internet voting, or simply \textit{i-voting}, has enabled voters to cast their ballots remotely, unlike those legacy voting systems which require voters to show up in a specific place like voting booths. Research shows that internet voting can be a better solution in pandemic periods, as it can give faster results and be more cost-effective when compared to mail voting \cite{pandemic} \cite{pandemic-internet1}. The i-voting brings possible advantages like reduced operational costs for elections, time-saving, increased voter participation, and improved transparency in elections. Estonia and Switzerland are two early adopters of \textit{i-voting} in nationwide government elections \cite{alvarez-hall-trechsel-2009}. An Estonian governmental agency, Enterprise Estonia (EAS), reported that the i-voting saved approximately 11,000 working days cumulatively in the 2011 Estonian parliamentary election \cite{noauthorestoniannodate}. The same report mentions that the saved costs were roughly equal to 504,000 euros. 

 Centralized nature of the Internet makes i-voting vulnerable to server-side attacks \cite{al-janabi-security-2019} and malpractice from central authorities such as censorship or modifying the election results \cite{p2p-censorship}. Blockchain has emerged as a new paradigm to solve these problems as it offers distributed, secure, privacy-preserving, and immutable applications. As reported in \cite{sym12081328}, the number of research publications on blockchain-based voting, {\it b-voting}, has increased from 7 to 30 in two years between 2017 and 2019. However, blockchain-based solutions also come with their own vulnerabilities
 
 All transactions are visible in public blockchains. Though the user addresses are pseudo-anonymous, data mining on the chains of transactions can leak various information about blockchain users. Using this information, collective and individual user behavior can be extracted  \cite{9364978}. This is of paramount importance in large-scale election systems because exposed voters can be manipulated with personalized messages prepared by online tools. Using permissioned blockchain does not alleviate the problem as it suffers from the trust in the authority \cite{10.1007/978-3-030-51280-4_3}.

Decentralization and trustlessness of public blockchain infrastructures are achieved by incentivized verifiers. This results in increased operational costs directly proportional to the number and the complexity of the transaction operations. Most of the b-voting proposals in the literature use complex encryption and decryption schemes to preserve vote privacy. Hence, they can only support small-scale elections with around ten to a hundred voters and one or a few candidates. 

Online voting prevents invalid votes, but some voters may abandon the election process at any stage. Both schemes require trust in the tallying bodies. Blockchain protocols make use of advanced cryptographic techniques to ensure vote privacy while preventing voter fraud such as duplicate voting or biased voting. Most of these techniques necessitate all registered voters to actively participate in the election. Otherwise, the election suspends forever or ends with no result. Hence, additional methods have been developed to handle voter abandonment but they are either too costly or inefficient in sustaining election security.

Based on these observations, this paper proposes ElectAnon which aims to improve the anonymity, robustness and scalability of blockchain-based election systems while satisfying all requirements for a secure election. Ranked-choice voting (RCV) is adopted since it is known to have a potential for more "democratized" elections \cite{anest2009ranked}. A candidate proposal system is also included in the protocol for an end-to-end decentralized election. ElectAnon is built on \textit{Ethereum Virtual Machine} (EVM) \cite{buterin2014next} so that it can support other EVM-based blockchain platforms such as Avalanche. It is implemented in such a way that the total election cost can be reduced up to 90\% of the existing b-voting proposals in the literature. This fact paves the path for a medium-scale (around 100,000 voters) election on blockchain at an acceptable cost. However, this is not an upper limit for ElectAnon: it can scale up indefinitely in the number of voters with increased costs. 

Our contributions can be listed as follows:
\begin{itemize}
    \item Secure election requirements are extended for b-voting: \textit{Anonymity} is derived from existing \textit{Privacy} requirements. The current definition of \textit{Robustness} is expanded by adding \textit{Autonomy} to it. \textit{Scalability} is  redefined and included in blockchain-based election requirements.
    \item Voters are represented by their individual identity commitments stored in a Merkle tree. The same tree is also used in generating zero-knowledge proofs to enable the voters to act anonymously in the elections while preventing them from duplicate voting and biased voting. The \textit{Merkle forest} extension is also proposed for adjusting trust assumptions and conducting even more trustless election environments.
    \item The protocol flows fully autonomously once the election authority manually starts the election. This is achieved by timeout parameters specified as the number of blocks between each stage. Even if there exist voters who fail to realize the duties required by the protocol, the election ends with the result obtained after tallying only committed votes. Invalid votes are not counted but kept as proof of commitment.
    \item Only the voters can interact with the protocol once voting starts and until the election completes. Each voter can interact several times with the protocol during the election to process and verify that her vote is committed, stored and tallied correctly. Biased voting is canceled by cryptographic identity commitment proofs generated from Merkle proofs. Hence, each voter can interact with only her vote but nobody else's vote. Intermediate results are eliminated by using cryptographic vote hashes in conjunction with encrypted votes. The election results are accessible by everybody, including the election authority, only after the election is complete.
    \item Solutions are proposed for scalability: Ranked-choice ballot is made independent of the number of candidates. Merkle trees are stored in batches to make the system scale up indefinitely in the number of voters at the cost of increased transactions. zk-SNARK proofs are used as they require less computation and storage compared to other types of zero-knowledge proofs. Voter registration is done by Election Authority to reduce voter expenses.
     \item Candidate proposal stage is made an integral part of the protocol to make the protocol also support decision-taking in decentralized autonomous organizations (DAO) where elections can be set up for hiring new employees, managing resources, deciding on feature sets, etc. 
     \item To make the protocol more flexible, a modular and algorithm-agnostic mechanism is used in tallying to switch between different methods easily. Two ranked-choice tallying methods, \textit{Borda Count} and \textit{Tideman}, are implemented and analyzed.
\end{itemize}

This paper is organized as follows: In the next section, secure election requirements are presented. Using them, the existing literature is analyzed. Subsequent four sections are reserved solely for ElectAnon: Section \ref{sec:prelim} gives preliminary information, Section \ref{system-model} explains the system model and protocol design, Section \ref{sec:options} describes the flexible implementation options, Section \ref{sec:electanon-security} analyzes the proposal under secure election requirements. Following them, the experimental results are presented. The final section concludes the work.
\section{Secure B-Voting Requirements}\label{election-requirement} 

The security of online elections has been studied by several works in the last twenty years. In 2002, D.A Gritzalis identified the requirements for a secure online election \cite{GRITZALIS2002539}. A recent work \cite{sensors} in 2021 mentions a similar set of requirements for secure blockchain-based online elections. This section presents an analysis of these requirements from the perspective of blockchain-based voting. Prominent works of the literature are also studied in light of these requirements to check whether they satisfy them. They are picked by studying a recent survey \cite{trends-in-blockchain} which comprehensively evaluates more than 50 b-voting proposals on different qualification criteria questions. Four highly-scored works are selected according to the following aspects: \begin{itemize}
    \item McCorry et al. (2017) (score=\%100). It is also one of the earliest b-voting proposals \cite{ovnet}.
    \item Chaintegrity (2019) (score= \%100). It has a smart contract implementation \cite{Zhang2019ChaintegrityBL}.
    \item Yang et al. (2020) (score= \%89). It uses a ranked-choice voting.  \cite{YANG2020859}. 
    \item Panja et al. (2020) (score= \%89). It uses a smart-contract based borda-count voting \cite{panja}.
\end{itemize}
We also included a more recent b-voting protocol, PriScore (2021), in the analysis as it considers \textit{score-voting} and has a smart contract implementation \cite{priscore}. These works are evaluated in the following aspects: \textit{Eligibility,  Uniqueness,  Privacy,  Universal Anonymity, Fairness,  Accuracy,  Universal  Verifiability,  Individual  Verifiability,  Robustness, Autonomy} and \textit{Scalability}. Table \ref{table:comparison} categorizes and summarizes this evaluation.

 Only eligible voters must cast ballots in an election \cite{GRITZALIS2002539}. Election owners and authorities generally decide a voter's {\bf \textit{eligibility}}. This requires trust in the authority as multiple voting accounts for the same identity can be made eligible for an election \cite{YANG2020859}. Even though the authority is trusted, eligibility is still a very fragile requirement in b-voting platforms because it may lead to vote-buying and coercion if not properly implemented. The voters need to know that they are eligible to commit a vote. In existing b-voting proposals, the blockchain addresses \cite{ovnet}\cite{panja} or the public keys of eligible voters \cite{Zhang2019ChaintegrityBL}, \cite{YANG2020859}, \cite{priscore} are used. 
Both approaches are vulnerable to \textit{linkage attacks} \cite{merener2012theoretical} as an anonymous digital identity can be linked to the actual identity. This vulnerability is further explained in "universal anonymity".

 The link between vote and voter should not be exposed \cite{Zhang2019ChaintegrityBL}.  This requirement has to be satisfied for all parties. {\bf \textit{Privacy}} is ensured in almost every b-voting proposal by blinding the votes with cryptographic techniques during vote commitment and tally phases. Yet, some studies may leak privacy when a voter commits a ballot but fails to participate in tallying his vote. When authorities, other voters or third parties step in to count the vote, the link between the abandoning voter and the vote is automatically established \cite{priscore}. 

The term "anonymity" and "privacy" is used interchangeably in the existing b-voting literature. It is observed that b-voting proposals claiming to be anonymous actually try to satisfy privacy. Based on this fact, we introduce {\bf \textit{universal anonymity}} as an extension of privacy: voting activities and blockchain activities should not be linked. Most blockchain addresses are pseudonymous by design \cite{https://doi.org/10.1002/nem.2130}. As a result, blockchain addresses can be traced to link the user with their chain activities. When voters act in an election (vote, commit, register, etc.), election authorities can know which addresses/public keys are used because voters need to prove that they are eligible to vote. Election authorities may form and enhance a link between the addresses/public keys and the vote usage pattern. This may expose a link between voter identities and blockchain activities such as transactions, token balances, and ownerships. Then, election authorities can link actual identities to these blockchain activities. This link can be further enhanced if the same list of eligible voters is used many times. Hence universal anonymity ensures that no parties, even election authorities, can know how the eligible voters interacted with the election protocol during the voting process. 

 Each voter should be able to cast at most one ballot \cite{GRITZALIS2002539} \cite{YANG2020859}.  Hence, {\bf \textit{uniqueness}} requirement eliminates double voting, i.e., a voter casts multiple ballots in an election. Privacy may break uniqueness if it is not carefully implemented. As privacy hides the relation between the vote and the voter, an eligible voter can vote multiple times by using a different voting key \cite{ovnet}\cite{panja}. It cannot even be detected when a clique of voters adjusts the number of their attempts by observing the number of non-participating voters. There exist successful implementations that bind voting keys with eligibility certificates. They do not allow the voting key to be used more than once.

No intermediate results should be available for all parties \cite{GRITZALIS2002539} \cite{sensors}. This requirement avoids biased voting: a voter can get influenced enough to change her intended choice. {\bf \textit{Fairness}} imposes two constraints: (1)  no vote should be revealed before tallying starts (2) tallying should start after the commit phase is over for all voters. Protocols that necessitate the keys of authorities or candidates while voting is susceptible to biased voting: it is possible for the corrupted authorities to organize an attack at any time for learning the intermediate results \cite{Zhang2019ChaintegrityBL}. Candidates may not share their secret keys in the tally phase if they are not satisfied with the ongoing election \cite{YANG2020859}. 

{\bf \textit{Accuracy}} requires only valid ballots to be considered in the tallying process.  It should not be possible to alter a vote in the final count. Invalid votes should be discarded and not tallied \cite{accuracy}.  Usually, zero-knowledge proofs or zero-knowledge range proofs are used to ensure the validity of a vote. Some protocols are set up in such a way that it is impossible to cast an invalid vote \cite{ovnet} \cite{panja}. Voters, who are unhappy with the choices may prefer not to vote in these systems. This may cause the election to suspend without a result as most blockchain systems rely on the commitments of all voters due to security concerns. 

Each voter should be able to verify that his ballot is cast correctly in the election \cite{priscore} \cite{Zhang2019ChaintegrityBL}. 
In b-voting literature, the voters observe their own blockchain transactions for {\bf \textit{individual verifiability.}}

The fairness and the correctness of an election result must be verifiable by anybody. Even non-participants must be able to validate the election result from the cast votes \cite{priscore}\cite{Zhang2019ChaintegrityBL}. Protocols that rely on only election authorities in tallying may not satisfy {\bf \textit{universal verifiability}}  \cite{YANG2020859}. {\bf \textit{Self-tallying}} seems to be a natural solution to this problem: everybody can tally the votes \cite{pkc-2002-3284}. It removes the need for authorities or specific actors to calculate the election result. 
This may lead to repeated elections where voters can get influenced in-between. Some protocols require additional input from other parties or voters so that one identity can count the votes \cite{YANG2020859} \cite{priscore}. We classify these studies as partially self-talliable. 

An election is robust if no party can disrupt an ongoing election \cite{sensors}. Shirazi et al. expand the {\bf \textit{robustness}}  by adding \textit{tallying availability} on top of \textit{voting availability} \cite{robustness}. Voting availability ensures that eligible voters can finish the voting process without disruption. Tallying availability ensures that valid votes can be tallied correctly without any interruption. Robustness is crucial for blockchain systems since each transaction requires a fee and cost. Disruption causes repeated elections, turning b-voting into a costly and inefficient voting platform. Some studies depend on third-party actors to decrypt ballots in tallying \cite{YANG2020859} \cite{Zhang2019ChaintegrityBL}. Yet, third-party actors can disrupt the election by providing invalid credentials or not providing credentials at all. There exist studies using shared secret-key encryption to encrypt the votes \cite{ovnet} \cite{panja} \cite{priscore}. Shared keys have to be revealed by all registered voters for tallying. If some voters abandon their votes and do not reveal their shared keys, the election will halt with no result. To solve this problem, either additional cryptographic operations \cite{priscore} or fund depositing/refunding mechanisms  \cite{ovnet} \cite{panja} have been proposed. Unfortunately, both methods have limited applicability. It has been shown that additional cryptographic operations do not work for more than one abandoned vote. Economic incentives can be surpassed by vote-buying or the political advantages of the election. 

Most internet-voting protocols typically consist of different phases like \textit{Initialization}, \textit{Voting} and \textit{Tallying} \cite{al-janabi-security-2019}. Fairness requires these phases to be non-overlapping. We define {\bf \textit{autonomy}}  as an extension of \textit{robustness}: there should be no halting/freezing between phase changes. Some studies require the election authority to make phase changes \cite{ovnet} \cite{panja}. These studies cannot be claimed to have autonomy as they rely on a party. Other studies do not mention how phase changes are realized. Studies failing to realize robustness cannot be considered autonomous as they can halt anytime. 

{\bf \textit{Scalability}} is the maximum number of voters and candidates in an election \cite{scalability}. In b-voting literature, implementations have been done either on abstract platforms or Ethereum, which aims to become a global and decentralized computer running many different applications called smart contracts \cite{buterin2014next}.  Like most public blockchain platforms, Ethereum has a limit on the block size, which also limits the number of transactions. This limit dynamically changes with the network state and the fork version. E-voting protocols are evaluated on Ethereum with the limits at the time of writing this paper \cite{etherscanio}. It has been observed that they either hit the maximum limit with a relatively small voter set \cite{ovnet}  \cite{priscore} or exceed the limit even with a single vote \cite{panja} \cite{YANG2020859}. Hence, they fail to realize large-scale elections. This high cost is mostly due to complex encryption/decryption schemes on the voter side to satisfy privacy requirements.  

\begin{table}
\caption{Evaluation summary of selected b-voting studies.}
\vspace*{5mm}
\centering
\resizebox{\columnwidth}{!}{%
\begin{tabular}{|c|c|c|c|c|c|c|c|c|c|c|c|c|c|c|}
\hline
\multirow{2}{*}{Work} &
  \multicolumn{9}{c|}{Election requirement} &
  Self-&
  \multirow{2}{*}{Scalable} &
 Electoral &
  \multirow{2}{*}{Platform} \\ \cline{2-10}
 &
  E &
  U &
  P &
  UA &
  F &
  A &
  UV &
  IV &
  \multicolumn{1}{c|}R &
  Tallying &
   & 
 System  &
   \\ \hline
McCorry et al.\cite{ovnet} &
  \checkmark &
  x &
  \checkmark &
  x          &
  \checkmark &
  \checkmark &
  \checkmark &
  \checkmark &
  o &
  \checkmark &
  \multicolumn{1}{c|}{x} &
  \multicolumn{1}{c|}{Yes-No} &
  \multicolumn{1}{c|}{Public/Ethereum} \\ \hline
  Chaintegrity \cite{Zhang2019ChaintegrityBL} &
  \checkmark &
  \checkmark &
  \checkmark &
  x &
  x          &
  \checkmark &
  \checkmark &
  \checkmark &
  x &
  x &
  \multicolumn{1}{c|}{x} &
  \multicolumn{1}{c|}{Choose-One} &
  \multicolumn{1}{c|}{Abstract} \\ \hline
  Yang et al. \cite{YANG2020859} &
  \checkmark &
  \checkmark &
  \checkmark &
  x          &
  x  &
  \checkmark &
  o &
  \checkmark &
  x &
  o &
  \multicolumn{1}{c|}{x} &
  \multicolumn{1}{c|}{Ranked Choice} &
  \multicolumn{1}{c|}{Abstract} \\ \hline
  Panja et al.\cite{panja} &
  \checkmark &
  x &
  \checkmark &
  x          &
  \checkmark &
  \checkmark &
  \checkmark &
  \checkmark &
  o &
  \checkmark &
  \multicolumn{1}{c|}{x} &
  \multicolumn{1}{c|}{Ranked-Choice} &
  \multicolumn{1}{c|}{Public/Ethereum} \\ \hline
  Priscore \cite{priscore} &
  \checkmark &
  \checkmark &
  o &
  x          &
  \checkmark &
  \checkmark &
  \checkmark &
  \checkmark &
  o &
  o &
  \multicolumn{1}{c|}{x} &
  \multicolumn{1}{c|}{Ranked-Choice} &
  \multicolumn{1}{c|}{Public/Ethereum} \\ \hline
  ElectAnon (this work) &
  \checkmark &
  \checkmark &
  \checkmark &
  \checkmark &
  \checkmark &
  \checkmark &
  \checkmark &
  \checkmark &
  \checkmark &
  \checkmark &
  \multicolumn{1}{c|}{\checkmark} &
  \multicolumn{1}{c|}{Ranked-Choice} &
  \multicolumn{1}{c|}{Public/Ethereum} \\ \hline
\end{tabular}%
}
\resizebox{\columnwidth}{!}{%
\begin{tabular}{c} 
\tiny E: \textit{Eligibility}, U: \textit{Uniqueness}, P: \textit{Privacy}, UA: \textit{Universal Anonymity}, F: \textit{Fairness}, A: \textit{Accuracy}, UV: \textit{Universal-Verifiability} IV: \textit{Individual-Verifiability}, R: \textit{Robustness} \\ 
\tiny \checkmark: implemented, x: not implemented, o: partially implemented
\end{tabular}
}
\label{table:comparison}
\end{table}

\section{Preliminaries}
\label{sec:prelim}

\subsection{Merkle Trees and Merkle Proofs}
A Merkle tree is a hash tree widely used in cryptography to swiftly verify the existence of a piece of data in a large unknown data set. It is a binary tree where the leaves contain the hashes for the data pieces in the data set. Intermediate nodes are obtained by hashing the siblings. Finally, the root node reflects the hash of all data pieces. The root hash is public or accessible by all parties that need Merkle proofs. 

Construction of the Merkle proof for a leaf starts with hashing its hash with the hash of the neighboring sibling. This process is repeated iteratively at each level up the tree until the root hash is obtained. The proof mechanism does not necessitate visiting all intermediate nodes. So, the path for obtaining the Merkle proof contains the hash of the leaf and the visited intermediate nodes. The size of this path is exactly the tree height, $height_{MT}$. Once the leaf and the proof path are given to the verifier, the time for verification is $O(height_{MT})$.

\subsection{zk-SNARKs and Semaphore}

Zero-Knowledge proofs provide a way to prove the existence of a certain information without revealing the information itself. They are used in anonymous authorizations, private payments, computation off-loading and b-voting \cite{morais2019survey}. \textit{Zero-Knowledge Succinct Non-Interactive Argument of Knowledge} or simply \textit{zk-SNARK} brought an efficient zero-knowledge protocol that reduces the number of rounds to verify proofs \cite{zksnark}. zk-SNARK proofs are "succinct" and can be verified in milliseconds, and their proof sizes can be as small as a few hundred bytes long. zk-SNARKs are widely used in blockchains to preserve privacy and off-load heavy computations for scalability. Ethereum optimized zk-SNARK operations and reduced their gas costs with the \textit{Istanbul} upgrade \cite{noauthor-istanbul-nodate}. 

zk-SNARKs are constructed with complex arithmetic equations called \textit{circuits} \cite{9300214}. Circuit compilers can abstract these equations and generate circuits using higher-level languages. Circom \cite{noauthor-circom-nodate} and Zokrates \cite{zokrates} are two popular zk-SNARK circuit compilers. They support basic software-language concepts like variables, functions, and control flows. These languages can declare private input signals which are not revealed in proofs. They can also generate smart contract verifiers to verify proofs on-chain. Typical steps for generating and verifying zk-SNARK proofs are listed as follows:
\begin{enumerate}    
    \item A high-level program (circuit) is designed by writing the logic for the zero-knowledge computation.
    \item The circuit is compiled into a set of low-level arithmetic equations.
    \item Verification and proving keys are generated with the compiled circuit.
    \item Prover executes the compiled circuit with given public and private inputs and generates the witness result.
    \item Prover generates the proof with the witness and the proving key.
    \item Prover sends the generated proof and public inputs to the verifier.
    \item Verifier verifies the proof with the verification key.
\end{enumerate}

In ElectAnon, \textit{Semaphore} \cite{semaphore-paper} is utilized as a zero-knowledge protocol for anonymous signaling. Semaphore aims to prove whether (1) an identity is eligible to broadcast a signal, (2) the signal truly belongs to the identity owner, and (3) the signal is broadcasted only once. It provides its users to prove these properties without revealing information about their identities. A valid proof verifies that the user is indeed on the eligible list. Eligible lists are defined with Merkle trees so that identity verification can be efficiently done with Merkle proofs. Semaphore contains a \textit{Circom} zk-SNARK circuit \cite{noauthor-circom-nodate} and two smart contracts. One of the smart contracts includes a Merkle tree implementation so that smart contract owners can register eligible identities to Merkle trees through the smart contract. The other smart contract verifies zero-knowledge proofs and prevents double-signaling by storing nullifiers. The project also provides a Javascript library to interact with smart contracts seamlessly and generate identities, witnesses, signals, and proofs. In ElectAnon, voter registration to the Merkle tree in Semaphore is modified to reduce the huge cost of voter registration.


\subsection{Ranked Choice Voting}\label{sec:permutation}

\textit{Ranked Choice Voting} (RCV) enables voters to sort their candidate preferences and vote with a sorted list, i.e., \textit{preference list}. Each preference in the list gives a ranked score to the respective candidate. James "Jim" Anest states the benefits of RCV for more "democratic" elections \cite{anest2009ranked} as follows: (1) RCV encourages voters to participate in elections because it enables voters to express their preferences without worrying about wasting their votes. (2) RCV can encourage candidates to enter elections without worrying about spoiling votes for other candidates. One of the debates claims that RCV could be too complicated for voters \cite{noauthor-ranked-choice-nodate}. However, Anest argues that advances in computer technology will facilitate employing RCV in large-scale elections \cite{anest2009ranked}.

Given $n_c$ candidates, there exist $n_c!$ possible formations, i.e. permutations, of a preference list. It is costly for each voter to form a private preference list when complex cryptographic operations including candidates' keys are required \cite{YANG2020859}\cite{priscore}. In ElectAnon, a voter's preference list is represented as a rank, i.e., an integer in the $[0...n_c!-1]$ range. Voters compute ranks of their preference lists off-chain (offline) on the smart contract by using Myrvold and Ruskey's $O(n_c)$-time algorithm to rank and unrank permutation lists \cite{permutations}. Privacy of the vote is achieved by encrypting only the rank. The same algorithm is also used in the reveal phase to unrank the vote before feeding it into the tallying algorithm. This approach dramatically reduces the gas costs on the voter side. 


\section{The ElectAnon Protocol Design}\label{system-model}
      
\begin{figure}[htbp]
	\begin{center}
		\includegraphics[trim=0cm 0cm 0cm 0cm, clip, width=1\textwidth]{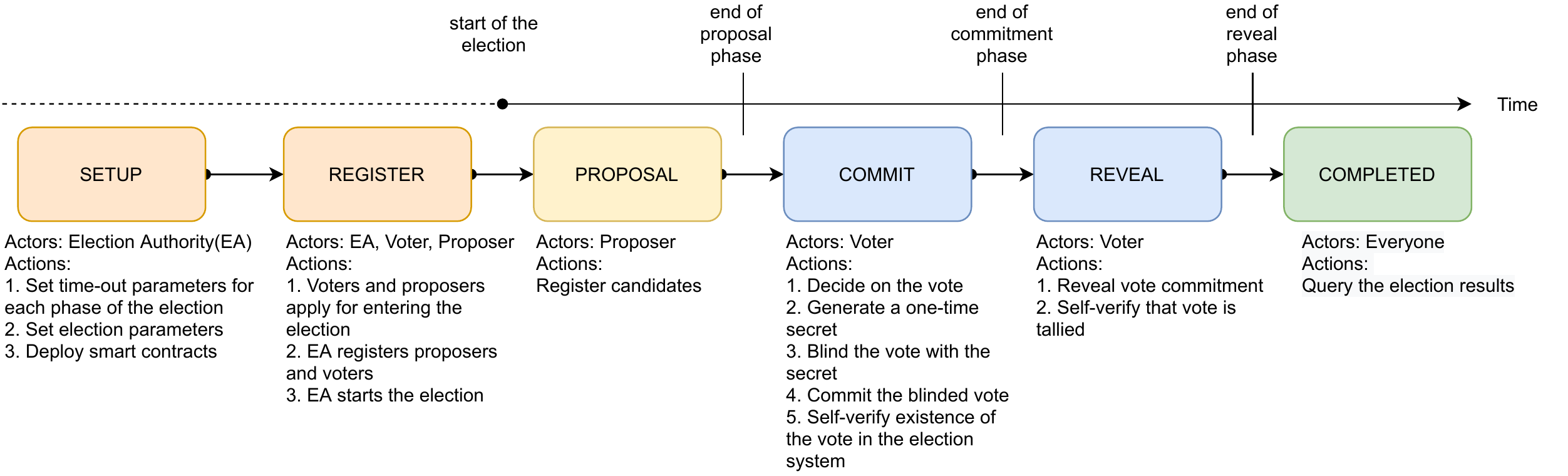}
		\caption{ElectAnon Protocol Flow}
		\label{fig:timeline}
		\Description[The timeline consists of Setup, Register, Proposal, Commit, Reveal and Completed phases]{In Setup phase there is only Election Authority actor. Election Authority sets phase timeout parameters, other election parameters and deploy smart contracts in this phase. In the register phase election authority, voters and proposers take role. Voters and proposers apply for the election, then election authority registers them in the election; and finally election authority starts the election. In proposal phase proposers register candidates. In commit phase voters decides their votes, generate one time secret, blind the vote with the secret and finally commit the vote; they can also verify that their commitment was successful. In reveal phase voters reveal their blinded votes and also verify their vote is tallied. In completed phase everyone can query the election results. Setup, Register are painted in light-red to indicate that EA takes a role in these phases. The timeline arrow is dashed in these phases since these phases can be manually changed. After the register phase the timeline arrow is continuous to indicate there is no interruption and manual change. The proposal phase is painted yellow to indicate proposers take a role. The commit and reveal phases are painted in blue to indicate voters take a role. The last phase is painted in green since everyone can take a role in this phase}
\end{center}
\end{figure}

In ElectAnon, there are three system actors: \textit{Election Authority (EA)}, \textit{Proposers} and \textit{Voters}. The protocol flows in a timeline of six phases as presented in Figure \ref{fig:timeline}. Different actors take different actions in each phase. Election Authority has a role only in \textit{Setup} and \textit{Register} phases. The timeline is dashed in these phases since these phases can be manually controlled by EA. Once EA starts the election, the timeline is solid to indicate there is no interruption and manual change. In \textit{Proposal} only the proposers can act. \textit{Commit} and \textit{Reveal} phases are open only to the actions of the voters. In the \textit{Completed} phase, the election results can be evaluated by everyone.

ElectAnon protocol implements two types of control mechanisms for the autonomous flow between \textit{Proposal}, \textit{Commit}, \textit{Reveal} and \textit{Completed} phases: (1) a timed-transition state machine (2) conditional transitions in functions to safely change the state to the next one. Each phase is represented by a state, which has a timed-transition guard with a relative deadline. If any state-specific transaction (function call) is received after this deadline, the transaction is rejected, and no changes occur in the blockchain. The relative deadline is given with respect to the start time of the phase in terms of the block numbers. If the actors of the current state finish their duties before the deadline, the phase change takes place before the deadline. For example, assume that the relative deadline of the Commit state is determined as 30 blocks. Then, any call to the Commit state-specific function is accepted if the two conditions are simultaneously satisfied: the call happens within the lifetime of 30 blocks after its start and there are still voters with uncommitted votes. If the block counter for Commit becomes 30 or all voters finish their commitments, the state changes to Reveal. Any call to the Commit is rejected after this time.

\subsection{Setup State}

\begin{figure}[htbp]
	\begin{center}
		\includegraphics[trim=0cm 0cm 0cm 0cm, clip, width=0.6\textwidth,]{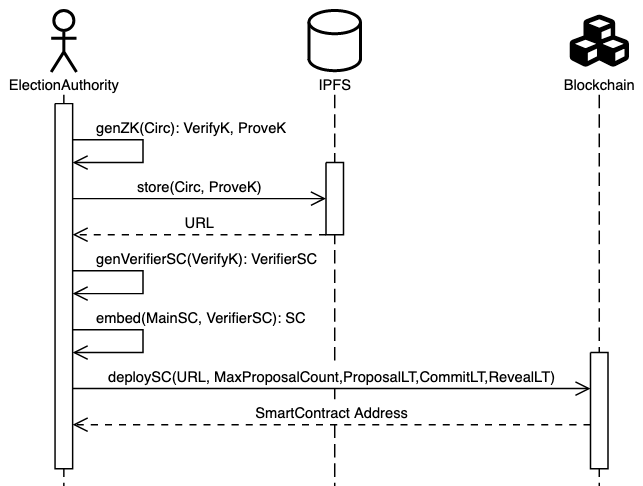}
		\caption{Setup State Sequence Diagram}
		\label{fig:setup-diagram}
			\end{center}
		\Description[The interaction between actors in the Setup State]{The sequence diagram shows the interaction between Election Authority, the InterPlanetary File System and Blockchain in the setup state}
\end{figure}

In this state, $EA$ has to decide on the election question. It can be an arbitrary string representing various question topics. For example, "who should be the president" in a presidential election and "which platform should be used" in organizational decisions. ($EA$) also initializes election parameters and prepares a zero-knowledge-proof environment. This phase ends with the deployment of smart contracts. Figure \ref{fig:setup-diagram} shows the sequence diagram for this state.

Election parameters are \textit{Merkle tree height, maximum proposal count, proposal lifetime, commit lifetime}, and \textit{reveal lifetime}. Merkle tree height, $height_{MT}$, determines the maximum number of voters as $2^{height_{MT}}$. The \textit{maximum proposal count} indicates the maximum number of proposals that can be registered as candidates. Each \textit{lifetime} ($LT$) parameter defines the lifetime of the related state in terms of block number.


$EA$ prepares the Semaphore zero-knowledge circuit ($\mathit{Circ}$). The circuit defines a static \textit{tree-level} parameter inside the circuit code to specify the maximum depth of the Merkle tree, $height_{MT}$. $EA$ can alter the tree-level parameter in the circuit to change the maximum voter count. Using Semaphore, $EA$ generates the zero-knowledge verification ($\mathit{VerifyK}$) and the proving keys ($\mathit{ProveK}$) with the given circuit $Circ$ as shown in the equation \ref{eqn:gen-zk}.
\begin{equation}
\label{eqn:gen-zk}
    gen^{ZK}(\mathit{Circ}) \xrightarrow{} (\mathit{VerifyK}, \mathit{ProveK})
\end{equation}
These keys have to be announced so that voters can verify the circuit and generate zero-knowledge proofs during the \textit{Commit} phase. EA can store these keys and the circuit in a place accessible by every voter. It can preferably be decentralized storage such as \textit{The InterPlanetary File System} (IPFS)\cite{ipfs}. The storage address ($URL$) can be put in the smart contract so that voters can fetch keys to generate their proofs.

$EA$ generates the verifier smart contract (\textit{VerifierSC}) with the verifier key ($\mathit{VerifyK}$). The verifier smart contract is embedded in the main smart contract (\textit{MainSC}) for a single smart contract ($SC$) deployment.

\subsection{Register State}

In this state, $EA$ starts forming eligible proposer and voter lists. Proposers send their blockchain addresses to $EA$ to propose in the election. Voters generate and send a commitment of their identities to $EA$. 

Voters generate their identity ($ID_{i}$) and identity commitment (${IDC}_i$). For the identity generation, $gen_{{ID}_i}^{EdDSA}(s_i)$ function takes a random seed $s_i$ and generates the $ID_i$ with the following components: a private key and public key pair ($PrivK_i,PubK_i$), a nullifier ($Null_{i}$) and a trapdoor ($\mathit{Trap_{i}}$). To reduce gas consumption, Edwards-curve Digital Signature Algorithm (EdDSA) \cite{semaphore-paper} is preferred in the generation of voter ID since it supports key aggregation. In this way, the voter will be able to sign the ballot at the \textit{Commit} phase in a single step. The identity commitment ${IDC}_i$ is constructed by Pedersen hash \cite{semaphore-paper} by hashing $PubK_i, Null_i$ and $Trap_i$ of $ID_{i}$.

\begin{align} 
\begin{split}
\label{eqn:gen-id}
    &gen_{{ID}_i}^{EdDSA}(s_i) \xrightarrow{} (PrivK_{i},PubK_{i},Null_{i},Trap_{i}): ID_{i}
    \\
    &h^{Pedersen}(PubK_{i},Null_{i},Trap_{i}) \xrightarrow{} {IDC}_i
\end{split}
\end{align}
Voters send their ${IDC}$s to $EA$ through a secure channel to get registered as eligible voters for the election. 
The identity $ID_{i}$ must be kept secret by the voter. 

\begin{figure}[htbp]
	\begin{center}
		\includegraphics[trim=0cm 0cm 0cm 0cm, clip, width=1\textwidth,]{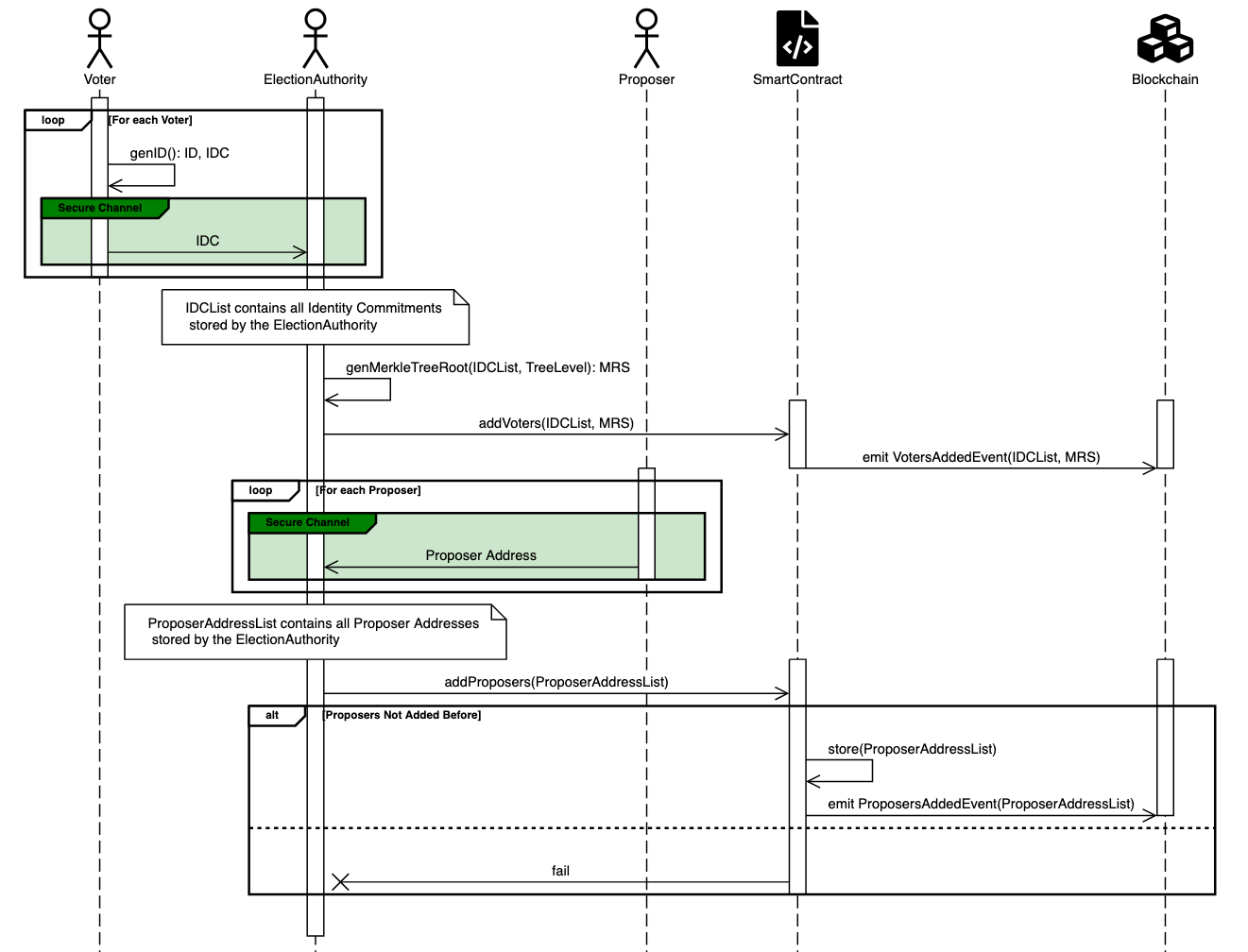}
		\caption{Register State Sequence Diagram}
		\label{fig:register-diagram}
			\end{center}
		\Description[The interaction between actors in the Register State]{The sequence diagram shows the interaction between Voter, Election Authority, Proposer, Smart Contract and Blockchain in the register state}
\end{figure}

$EA$ generates a Merkle root ${MR}_S$ from all collected identity commitments, then submits ${MR}_S$ along with the list of collected $IDC$s to the smart contract. The smart contract has a function, \textit{addVoters}, for Merkle tree registration. The function takes the ${IDC}$ list and the Merkle root ${MR}_S$. Due to the gas limit of Ethereum, ${IDC}$ list size has a limit. In our experiments, this limit is found to be around 30.000 $IDC$s for a single \textit{addVoters} call. $EA$ can split the ${IDC}$ list into smaller batches (like 30.000 per call) and issue them with multiple calls to the smart contract. ${IDC}$s must be added to the smart contract in the same order as they are used when constructing the Merkle root. This enables the voters to reconstruct the same Merkle tree that will be used during zero-knowledge proof generation at the \textit{Commit} phase. When registering is finished, $EA$ can manually change the state machine to the \textit{Proposal} state. Figure \ref{fig:register-diagram} shows the sequence diagram for this state.

\subsection{Proposal State}

\begin{figure}[h!]
	\begin{center}
	\includegraphics[trim=0cm 0cm 0cm 0cm, clip, width=0.6\textwidth,]{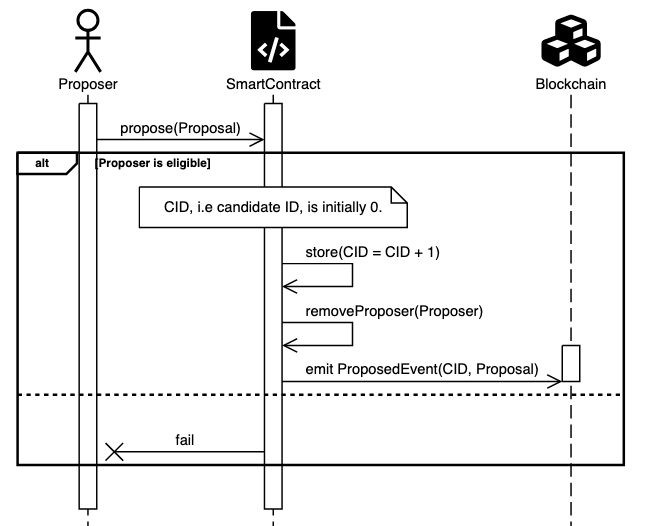}
	\caption{Proposal State Sequence Diagram}
	\Description[The interaction between actors in the Proposal State]{The sequence diagram shows the interaction between Proposer, Smart Contract and Blockchain in the proposal state}
	\label{fig:proposal-diagram}
		\end{center}
\end{figure}

Proposers can send their proposals to the smart contract ($SC$), which assigns a \textit{candidate ID} ($CID_i$) for each proposal. Proposals can be considered as the proposed answer strings for a possible solution to the underlying election question. Voters use $CID$s in their preference lists. Proposals are announced and stored in the blockchain as event logs. The smart contract removes the proposers from the eligible list after proposing to ensure they can only submit once. The smart contract publishes a \textit{ProposedEvent}, which contains the assigned candidate ID ($CID_i$) and the proposal string. Note that the number of candidates can be less than the maximum candidate count: $n_c \leq \max_{i}CID_i$+1. $EA$ can choose to register more than the maximum candidate count proposers in the \textit{Register} state to increase the chance of having enough proposals for the election. Figure \ref{fig:proposal-diagram} shows the sequence diagram for this state. Transition to the next state is autonomously done either at the end of the proposal lifetime or  at the time when $n_c =$ \textit{maximum candidate count}. 

\subsection{Commit State}

Voters cast their ballots in the \textit{Commit} state with a new blockchain address to ensure \textit{Universal Anonymity}.  Each voter decides its preference list after studying the candidate list fetched from the blockchain by filtering \textit{ProposedEvent} events. The preference list must reflect the voter's choice from the most preferred to the least preferred. The list has to contain every registered candidate. Each preference lists is a permutation of $[1,2,3,...,n_c]$. ElectAnon uses the ranking algorithm (Section \ref{sec:permutation}) to map a given permutation list to a single rank integer. The voter uses this rank as her ballot, i.e. the vote ID (${VID}_i$). 

ElectAnon uses vote hashes ($VH$s) to hide the plaintext ${VID}$s to preserve \textit{Fairness}. The voter can hash his ${VID}_i$ with a secret key $\mathit{VSk}_{i}$. In order to mitigate an attack, $\mathit{VSk}_{i}$ should be a big random number. \textit{Keccak-256} is used for hashing since it is a gas-efficient cryptographic hash function. 
Both ${VID}_i$ and ${VSk}_i$ should be kept secret to preserve privacy.
\begin{equation}
\label{eqn:vh}
    h^{keccak256}(\mathit{VID}_{i}, \mathit{{VSk}_i}) \xrightarrow{} \mathit{VH}_i
\end{equation}

Voters generate Merkle trees to obtain their Merkle proofs to prove their eligibility to vote. Merkle tree leaves ($IDC$s) can be fetched from blockchain by filtering \textit{VotersAddedEvent} event logs. Voters also obtain the Merkle tree-level from the smart contract. Then, each voter can generate the Merkle tree and the Merkle proof $MP_i$ with her own $IDC_i$. The equation \ref{eqn:genTree} shows how to generate the Merkle proof.
\begin{align} 
\begin{split}
\label{eqn:genTree}
    &genTree(IDCList, TreeLevel) \xrightarrow{} MerkleTree
    \\
    &genMerkleProof(MerkleTree, {IDC}_i) \xrightarrow{} MerkleProof(MP_i)
\end{split}
\end{align} 
 
Then, each voter can start generating the witness and the zero-knowledge proof ($P_i$) with the \textit{Semaphore} circuit ($Circ$). She can fetch the circuit ($Circ$) and the proving key ($ProveK$) from the \textit{IPFS} with the $URL$ provided in the smart contract. The circuit expects the following inputs: the vote hash (${VH}_i$), the identity commitment (${ID}_i$), the Merkle proof $MP_i$, the external nullifier ($ExtN$), and the signature ($Sign_i$). The $ExtN$ is the same as the contract address and accessible from the smart contract. It acts as the voting booth, i.e., anonymous signaling is done to $SC$. $Sign_i$ is obtained by signing the $VH_i$ with the private key $PrivK$. The circuit outputs the witness, which contains a verification for the Merkle root ${MR}_C$ and the nullifier hash $NH_i$. A detailed explanation about ${MR}_C$ and $NH_i$ is given at the end of this section. Voters can use this witness result and the $ProveK$ to generate their proofs $P_i$.
\begin{align} 
\begin{split}
\label{eqn:genProof}
    &genWitness(Circ, VH_i, ID_i, MerkleProof, ExtN, Sign_i) \xrightarrow{} Witness: MR_C, NH_i
    \\
    &genProof(Witness, ProveK) \xrightarrow{} P_i
\end{split}
\end{align} 

\begin{figure}[hbpt]
	\begin{center}
		\includegraphics[trim=0cm 1cm 0cm 0cm, clip, width=0.75\textwidth,]{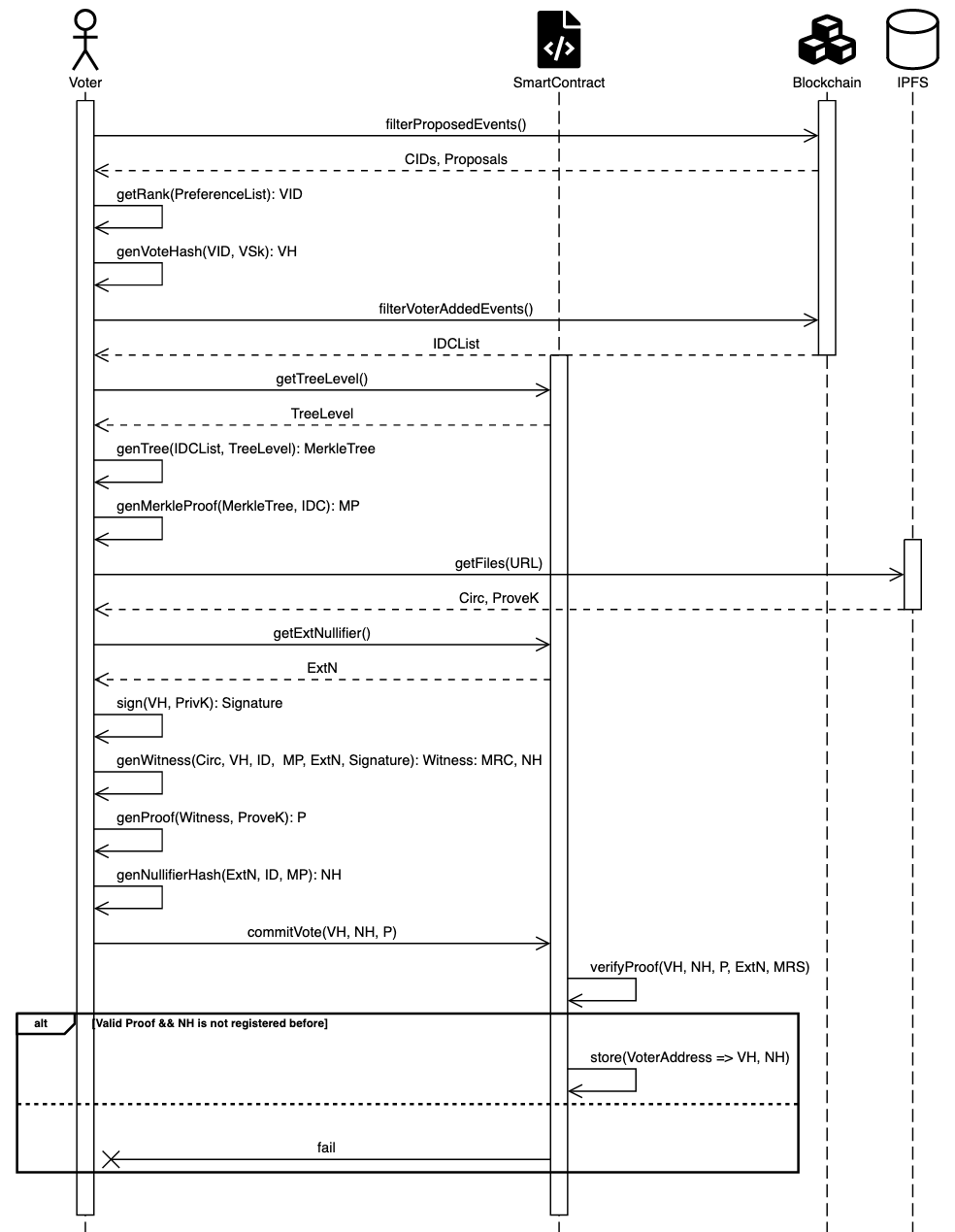}
        \caption{Commit State Sequence Diagram}
    	\Description[The interaction between actors in the Commit State]{The sequence diagram shows the interaction between Voter, Smart Contract and Blockchain and IPFS in the commit state}
        \label{fig:commit-diagram}
        	\end{center}
\end{figure}

The \textit{commitVote} function in the smart contract expects three inputs from voters: ${VH}_i$, $NH_i$ and $P_i$. The verifier smart contract (\textit{VerifSC}) expects two more inputs to verify $P_i$: $ExtN$ and ${MR}_S$. $ExtN$ is defined as the contracts deployed blockchain address. ${MR}_S$ is the registered Merkle root in the \textit{Register} state. Then, the smart contract verifies the given proof $P_i$ to ensure the proof is intact with the given inputs. The contract either rejects or accepts the proof. If the proof is accepted, \textit{nullifier hash} ($NH_i$) is marked as used to prevent the double-voting. The smart contract stores the vote hash $\mathit{VH}_i$ in a mapping keyed with the voter's blockchain address for later use. Figure \ref{fig:commit-diagram} shows the sequence diagram for this state.

The \textit{Commit} state ends with two conditions. The first one is the timed transition which changes the state after \textit{commitLifetime}. The smart contract will reject any transaction (function call) made after this time. The other one checks whether every voter has already committed.

{\bf \textit{The Semaphore Proof \textemdash}}
It is worth mentioning how the Semaphore proof $P_i$ is constructed and works in our voting protocol. In general, the proof ensures these three properties:
\begin{enumerate}
    \item The voter identity is in the eligible voter list. 
    \item The same identity is not used to cast a vote hash twice.
    \item The vote hash is indeed generated by the identity which created the proof.
\end{enumerate}

The circuit guarantees the first property with Merkle trees and Merkle proofs. Merkle proof ($MP_i$) of the voter is used as a private input in the circuit. The circuit also takes each component of ${ID}_i$ as private inputs. 
The circuit re-generates the ${IDC}_i$ by hashing $PubK_i$, $Null_i$ and $Trap_i$ as in Equation \ref{eqn:gen-id}. The circuit generates the Merkle root $MR_C$ with the generated ${IDC}_i$ along with the given $MP_i$. The circuit puts the Merkle root ${MR}_C$ into the proof $P_i$. The smart contract verifies $P_i$ and checks whether the root in the proof ${MR}_C$ is verifiable with the registered root ${MR}_S$. The verifier verifies that the voter has indeed generated the correct Merkle root with his own ${IDC}_i$ and $MP_i$. Hence, ElectAnon verifies that the voter is eligible as the ${IDC}_i$ is indeed a member of the eligible voter Merkle tree.

The circuit ensures the second property with a nullifier hash ($NH_i$). The circuit takes components of $ID_i$ as private inputs. One of them is \textit{identity nullifier} (${Null}_i$), which is a random integer. The circuit hashes following inputs to obtain $NH_i$: $Null_i$, $ExtN$ and $MP_i$. Semaphore uses \textit{Blake2s} for this purpose since it is safely
used in place of a random oracle in many cryptographic applications. 
\begin{equation}
    h^{Blake2s}(Null_{i}, ExtN, {MP}_i) \xrightarrow{} NH_{i}
\end{equation}
The circuit puts $NH_i$ into the proof $P_i$. In the verification, the smart contract takes $NH_i$ from the voter and verifies that it matches the one in the $P_i$. Additionally, the smart contract marks and stores this $NH_i$ as used and invalidates any future calls with the same $NH_i$. This prevents double voting with the same $NH_i$. The zero-knowledge proof ensures that $NH_i$ is constructed correctly with components of $Null_{i}, ExtN, {MP}_i$. So reforging a new $NH_i$ would require a change in these components. $ExtN$ is provided by the smart contract itself to the verifier, so the voter has no direct control over it. If the voter provides an invalid $ExtN$ in the proof generation, the verifier will not validate the proof. $Null_i$ and $MP_i$ is a part of ${IDC}_i$, so reforging $Null_i$ would result in a completely different ${IDC}_i'$. This would invalidate the proof $P_i$ as this new ${IDC}_i$ would not be in the eligible list. 

Voters sign their vote hashes ($\mathit{VH}_i$) with their private key $PrivK_i$. The circuit takes the voter public key $PubK_i$ and the signature $Sign_i$ as private inputs. Then, it checks the signature $Sign_i$ with the given public key $PubK_i$. This completes the last property since it can verify that the voter indeed generates $VH_i$.
 
\subsection{Reveal State}

\begin{figure}[h!]
	\begin{center}
		\includegraphics[trim=0cm 0cm 0cm 0cm, clip, width=0.6\textwidth,]{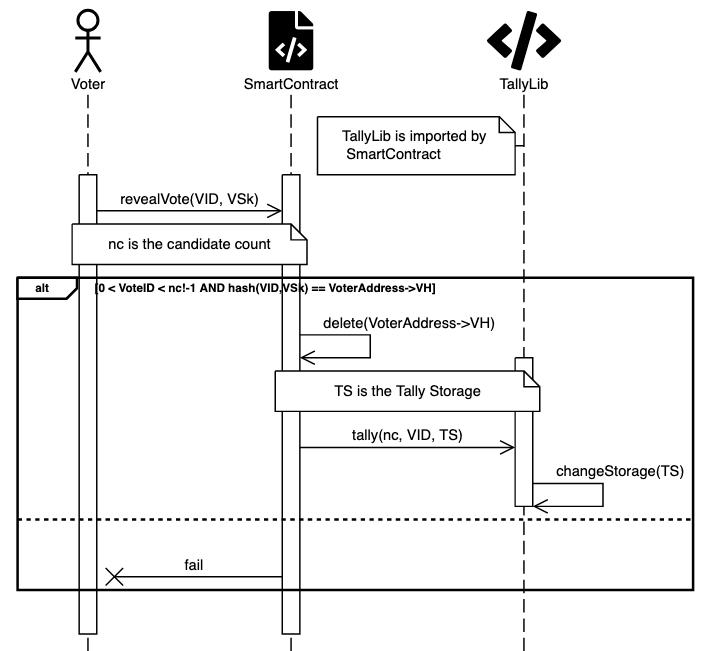}
    	\Description[The interaction between actors in the Reveal State]{The sequence diagram shows the interaction between Voter, Smart Contract and TallyLib in the reveal state}
		\caption{Reveal State Sequence Diagram}
		\label{fig:reveal-diagram}
			\end{center}
\end{figure}

Each voter reveals her vote hash $\mathit{VH}_i$ with \textit{the revealVote} function. The contract stores $\mathit{VH}_i$ in the \textit{Commit} state within a map of \textit{addresses} to \textit{vote hashes}. Hence, voters have to reveal their commitments with the same blockchain addresses they used in the \textit{Commit} state. Each voter provides \textit{voteID} ($\mathit{VID}_i$) and the vote secret key $\mathit{VSk}_i$ to the smart contract. The smart contract checks if the hash of these two inputs ($\mathit{VH}_i'$) is equal to the one stored ($\mathit{VH}_i$) in the previous state. If they're not equal, the smart contract rejects the transaction. If hash holds, the smart contract deletes the stored $\mathit{VH}_i$ in the contract to guarantee that it is not revealed twice. 
The contract passes the $VID_i$, the candidate count ($n_c$), and the storage mapping \textit{Tally Storage} ($TS$) to the tally library. $TS$ is required to keep revealed votes in the storage so that tally libraries can use revealed votes. The tally library defines a \textit{tally} function that stores revealed results in $TS$. \textit{tally} function can be customized for different algorithms. Figure \ref{fig:reveal-diagram} shows the sequence diagram for this state. The state is changed if the timed-transition \textit{revealLifetime} exceeds or all committed votes are successfully revealed. Voters must reveal their votes within this lifetime; otherwise, they cannot be counted.

\subsection{Completed State}

\begin{figure}[h!]
	\begin{center}
		\includegraphics[trim=0cm 0cm 0cm 0cm, clip, width=0.6\textwidth,]{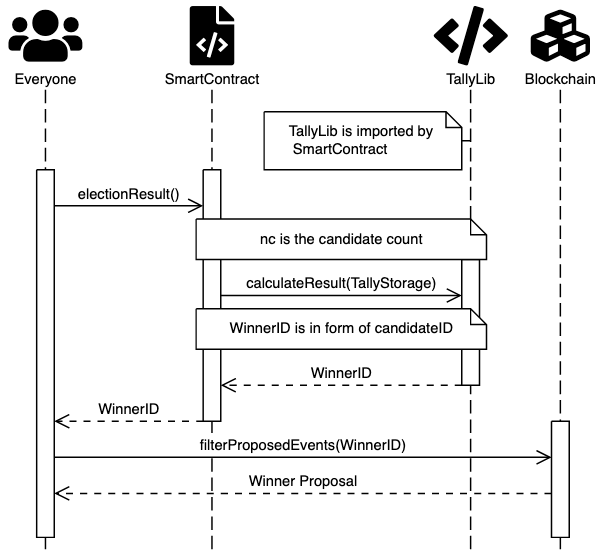}
		\Description[The interaction between actors in the Completed State]{The sequence diagram shows the interaction between Everyone, Smart Contract, TallyLib and Blockchain in the completed state}
        \caption{Completed State Sequence Diagram}
        \label{fig:completed-diagram}
        	\end{center}
\end{figure}

Everyone can call the \textit{electionResult} function to get the election result in this state. This function is a \textit{view} function, meaning that calling it will not create state-changing transactions in the blockchain and requires no transaction fees. The smart contract uses a tallying library to calculate the results. The tallying library uses the tallying storage $TS$ and calculates the election result. $TS$ is already populated in the \textit{Reveal} state with revealed votes. The \textit{electionResult} interprets the votes in $TS$ and shows the candidate ID (${CID}_i$) of the winner. As the smart contract has already published \textit{ProposedEvent} for each proposal in the \textit{Proposal} state, the actual proposal string can be fetched by filtering \textit{ProposedEvent} with the winner ${CID}_i$ to find the winner proposal. Figure \ref{fig:completed-diagram} shows the sequence diagram for this state.

\section{Implementation Options}
\label{sec:options}
\subsection{Tallying Libraries}\label{sec:tally-lib}
ElectAnon is designed to support different tallying libraries. The tallying library which is adopted in an election should be selected in the \textit{Setup} phase. 
A tallying library should implement two main functions, \textit{tally} and \textit{calculateResult}, as follows: The \textit{tally} function is used in the \textit{Reveal} state. It counts the votes and puts them into the tallying storage $TS$. The \textit{calculateResult} function is used in the \textit{Completed} state. It interprets the given tallying storage $TS$ and announces the winner candidate ID, i.e. CID. Currently, two tallying libraries are implemented in ElectAnon: Borda and Tideman. 


The Borda Count was devised by Jean Charles de Borda in 1781 \cite{emerson-original-2013}. In the basic form, each ballot holds a sorted list of candidates. Each of these candidates is assigned a score based on their orders in the list. At the end of the election, all scores for each CID is summed. The candidate with the maximum total score wins the election. In the Borda Count library of ElectAnon, the \textit{tally} function takes voteID ($VID$) and then unranks it into the related preference list. The preference list represents a sorted list of candidates. Then, each of the sorted candidates is scored in decreasing values as $[n_c, n_c-1, n_c-2,...,1]$, where $n_c$ is the candidate count. These scores are added to the tally storage $TS$ in a map of CIDs to their respective cumulative score. \textit{calculateResult} function compares each cumulative score of candidates and then returns the CID with the maximum score. 

The Tideman method was proposed by T. N. Tideman in 1987 \cite{tideman}. It collects the ranked preferences and compares each candidate in a pairwise fashion. The pairwise comparisons are sorted by their winner's vote dominance against the loser. The algorithm starts locking the winners against losers in this sorted order by constructing a directed graph. The one that is not locked by another candidate becomes the winner. If any cycle occurs in the locking, that pair is ignored. 
Unlike Borda Count, Tideman method supports both \textit{majority rule} and \textit{condorcet} criterion \cite{Homp2019ContemporaryMC} \cite{tideman}. In the Tideman library of ElectAnon, the \textit{tally} function uses a counter for each rank. Each counter-rank pair is stored in the tally storage $TS$
The \textit{calculateResult} function obtains preference lists from the ranks and follows the Tideman algorithm to calculate the winner. Both of algorithm details and pseudocode can be found under \ref{chapter:tallying-algorithms} in Appendices.

\subsection{Merkle Forest}

ElectAnon uses Merkle trees to prove eligibility of voter identity commitments efficiently. The original Semaphore smart contract calculates the Merkle tree on the smart contract (on-chain) to provide a safe way to keep the Merkle root intact with the registered leaves. However, it causes a very high gas consumption due to the hash operation at each intermediate node.  ElectAnon offloads the calculation to the election authority (EA) to eliminate gas consumption. This approach requires trust in the EA: he has to construct the Merkle tree honestly. 

To reduce trust in the EA, ElectAnon proposes calculation of the Merkle root with a zero-knowledge circuit and verified on the chain. Note that ElectAnon already has a zero-knowledge proof construction from the Semaphore circuit. This additional solution introduces another zero-knowledge circuit and verifier to the protocol to verify that Merkle tree constructions are done faithfully. In this approach, the circuit takes a fixed-size leaf-list as a private input. Then it forms a Merkle tree from these leaves and generates the Merkle root. It also hashes all elements in the given list input to obtain the hash output. The smart contract takes the leaf-list and the zero-knowledge proof, which is made up of the hash output and the Merkle root. Then, it computes the hash of the given leaf-list to verify the given zero-knowledge proof. In this way, the smart contract can ensure that the given Merkle root is indeed calculated with the given leaf-list. After a successful verification, the root is registered to the ElectAnon smart contract. 

zk-SNARK circuits do not support dynamic-size arrays. For instance, if the fixed tree size in the circuit is set to 256, each tree can hold at most 256 voters. In the Merkle forest solution, the voters are distributed in different Merkle trees of the same height. In order to register 2560 voters for the election, ten trees of size 256 constitute the Merkle forest. Smart contract stores tree roots within a \textit{mapping[treeIndex] => treeRoot} structure. It publishes an event when a tree is registered. The event contains \textit{treeIndex, treeRoot}, and \textit{leaves} so that voters can track which tree they're registered. In the \textit{Commit} state, each voter must specify their tree index; so that the smart contract can fetch the Merkle root of that particular tree and verify the Semaphore proof of the voter. With this method, the smart contract can verify that each Merkle root input is correctly constructed from the given Merkle tree leaves. 

Semaphore uses the \textit{Keccak-256} hash function for generating the hash output. It is an inefficient function for zero-knowledge proofs, but an efficient hash function for EVM-based smart contracts. Merkle forest solution requires Keccak-256 only when leaf inputs are private. If the leaf-list is made public, the hash function can be removed from the circuit. Then, the verifier smart contract can take the complete leaf-list as the input and verify the remaining Merkle tree construction proof with this list input. 

\section{Security Analysis}
\label{sec:electanon-security}

ElectAnon is designed for communities where the voters value their votes and care about their anonymity. The protocol is tolerable to misbehaving participants. If a voter deliberately discloses his vote or his credentials, this does not harm the privacy or anonymity of the other voters. If a voter abandons his vote, the election is not abandoned. As only the voters can interact with the system during voting and tallying, other participants cannot manipulate the votes. During the interaction, each voter acts on his behalf. Hence, no voter can see how the other voted. Besides, ElectAnon is not disruptable once the election starts. Misbehaving EA and proposer can cause ElectAnon to end with no winner at no cost to the voter but cost to the EA: (1)EA does not post the IDC list in the order used in Merkle tree generation. In that case, voters cannot generate the Merkle root correctly to start vote commitment. (2) Proposers do not propose within \textit{proposal lifetime}. In that case, voters cannot form their ballots to start vote commitment. Further analysis of ElectAnon with respect to the requirements defined in Section 2 is presented in the rest of this section.

ElectAnon ensures {\bf \textit{eligibility}} as follows: At the \textit{Register} state, each voter generates an identity commitment ${IDC}_i$ and submits it to the election authority ($EA$). $EA$ decides voters' eligibility and forms a Merkle tree ($MT$) with these ${IDC}$s. $EA$ registers the Merkle root (${MR}_S$) and the identity commitment list (${IDC}$s) to the smart contract. At the \textit{Commit} state, each voter creates proofs to prove that he owns a valid identity commitment ${IDC_i}$. Voters can generate the correct Merkle root with their Merkle proofs $MP_i$. The smart contract verifies this proof with the registered ${MR}_S$. At the end of this verification, the protocol verifies that the ${IDC}_i$ is indeed in the eligible voter set. The smart contract rejects all transactions lacking valid proofs. 

{\bf \textit{Privacy}} is assured by preserving the anonymity of voter identities. Voters should send their identity commitments (${IDC}_i$) to \textit{Election Authority} ($EA$) in a secure, private channel. Then, the protocol ensures that identities $ID_i$ and their commitments ${IDC}_i$ are secret at the \textit{Commit} state as follows: Voters generate zero-knowledge proofs without revealing their identity commitments ${ID}_i$ or ${IDC}_i$. The zero-knowledge proof $P_{i}$ proves that the vote is indeed generated by the voter. Voters generate their zero-knowledge proofs in their local offline environments. The zero-knowledge circuit $Circ$ takes voter identity ${ID}_i$ as a private input and generates the proof $P_i$. The zero-knowledge compiler ensures that the given private input of ${ID}_i$ is not revealed in the $P_i$. The smart contract takes $P_i$ as input but does not require ${ID}_i$ or ${IDC}_i$. As a result, no identity $ID_i$ or their commitment counterpart ${IDC}_i$ is revealed.  A privacy breach is possible if and only if the voter deliberately reveals his ${ID}_i$ or ${IDC}_i$ at this state. Then $EA$ can distinguish voters and learn their votes. 
For {\bf \textit{universal anonymity}}, each voter must also use a new and fresh blockchain address when he first interacts with the smart contract at the \textit{Commit} state because interaction with the smart contract is done through the blockchain addresses. Any reuse of these addresses in other blockchain applications can compromise universal anonymity. 

Semaphore is utilized to prevent double-signaling \cite{semaphore-paper}. Voter identity $ID_{i}$ includes a random-secret nullifier $Null_{i}$. Each voter generates a zero-knowledge proof $P_{i}$ with private inputs of  his nullifier $Null_{i}$ and an external nullifier $ExtN$. The smart contract verifies that $NH_{i}$ is correctly generated by the voter with the $P_i$. The smart contract also stores $NH_{i}$ and invalidates any call with the same $NH_{i}$. Since each $NH_{i}$ is uniquely generated from ${ID}_i$, one must regenerate a new $ID_i$ to reforge a valid $NH_{i}$. If the new $ID_i$ is not registered in the eligible set, the proof will still fail due to the \textit{Eligibility} proof. So each voter can only vote once with one valid $ID_i$. Thus, {\bf \textit{uniqueness }} is ensured in ElectAnon.

Voting in ElectAnon consists of two consecutive states, \textit{Commit} and \textit{Reveal} to ensure {\bf \textit{fairness}}. Voters commit the hash of the vote (${VH}_i$) without revealing the actual vote ID (${VID}_i$) at the \textit{Commit} state. This is done by hashing ${VID}_i$ with a secret vote key ($\mathit{VSk}_i$). The votes are not modifiable once they are committed since they are stored on the smart contract and $NH_i$ prohibits double voting. Hence, the protocol ensures \textit{Fairness}. It can leak if and only if some voters deliberately announce their ${VID}_i$ or $\mathit{VSk}_i$ before the end of the \textit{Commit} state.

At the \textit{Reveal} state, voters provide their plaintext ${VID}_i$ along with ${VSk}_i$ as their inputs to the smart contract. The smart contract verifies that committed hashes can be reconstructed with these inputs. The revealed votes are stored in the smart contract to be tallied by the smart contract itself. This ensures that ${VID_i}$s are not modifiable in the final count. ElectAnon encodes ranked-choice lists into a single integer, i.e. rank. Each rank represents a permutation of candidate IDs. For a list of size $n_c$, there can be at most $n_c!$ permutations. This means that a vote is valid if and only if ${VID_i}$ is in the range of $[0, n_c!-1]$. At the \textit{Reveal} state, ElectAnon checks the revealed ${VID}_i$ and rejects the transaction if it is not in the valid range. Hence only valid ballots are tallied. It is impossible to change the committed vote as the smart contract also secures the revealed votes. Thus, ElectAnon satisfies the {\bf \textit{accuracy}} requirement.


Voters can verify that their committed vote hashes ($VH$s) are successfully cast in the ballot by verifying blockchain transaction inputs. They can also ensure that their revealed votes are tallied correctly by verifying that transactions contain their plaintext ($VID$) inputs. Thus ElectAnon achieves {\bf\textit{individual verifiability}}.

Blockchain technology ensures that every transaction is transparent and verifiable by the whole network. ElectAnon uses a \textit{self-tallying} protocol so that everyone can calculate results independently. It achieves self-tallying with the \textit{tally} and \textit{calculateResults} functions. Voters use \textit{tally} function to reveal their vote preferences. The smart contract stores revealed votes in the storage called \textit{tallying storage} $TS$, which is a part of the smart contract. After this point, voters' preferences are available to public. \textit{calculateResults} function interprets this $TS$ accordingly to the selected tally library and returns the winner. This function can be called in the \textit{Completed} state by everyone. With these two aspects, ElectAnon achieves {\bf \textit{universal verifiability}}.

A timed-transition state machine is implemented in the smart contract to ensure {\bf \textit{autonomy}}. So election protocol can change states without interruption. The \textit{Register} is the only state that the protocol requires a manual transition from the election authority ($EA$).  $EA$ can disrupt the protocol before the \textit{Register} state, but it will not affect voters because they cannot interact with the smart contract before the \textit{Commit} state. Hence, ElectAnon guarantees that a started election will not be disrupted by any means. As a result, ElectAnon achieves \textit{autonomy}. Voters' activities also do not affect each other. Each voter prepares her vote locally and submits it to the correct smart contract. Hence, \textit{voting availability} is also ensured. In the \textit{Reveal} state, voters reveal their votes. If a voter abandons a committed vote, others will not be affected by it and the protocol can safely continue to tally. As ElectAnon is a self-tallying protocol, everyone can tally the result without requiring any external assistance. Hence, ElectAnon also satisfies the \textit{tallying availability}. Thus, ElectAnon ensures {\bf \textit{robustness}} by combining \textit{voting availability}, \textit{tallying availability}, and \textit{autonomy}.

ElectAnon is {\bf \textit{scalable}} indefinitely in terms of the number of voters. This is achieved by  keeping voter costs independent of voter count. There is no registration cost for the voter since registration is done by the EA. Each voter acts by himself: he encodes the ballot into a single integer and blinds the vote with zk-SNARK proofs off-chain. After he submits the vote, the verification of the proof is done on-chain. Hence, interaction of each voter with ElectAnon consumes a constant gas which has to be paid by the voter. 


\section{Experimental Evaluation}\label{sec:experiments}

ElectAnon is deployed to a local Ethereum network with the latest Ethereum fork \textit{London} by using the Hardhat \cite{hardhat} tool, which is a development environment that provides local Ethereum networks, gas consumption reports, a high-level language to conduct tests, and a wallet pre-filled with accounts. A modified version of Semaphore library \textit{libsemaphore} \cite{libsemaphore} is utilized to generate proofs and witnesses.  Smart contracts are developed with Solidity version 0.8.7. Tests in NodeJS and executed on a MacBook Pro with an 8-core 3.2GHz Apple M1 chip and 16 GB Ram, macOS version 11. The tallying library is selected as Borda Count. 

\begin{figure}[htbp]
	\begin{center}
		\includegraphics[trim=1cm 17cm 0cm 0cm, clip, width=0.85\textwidth,]{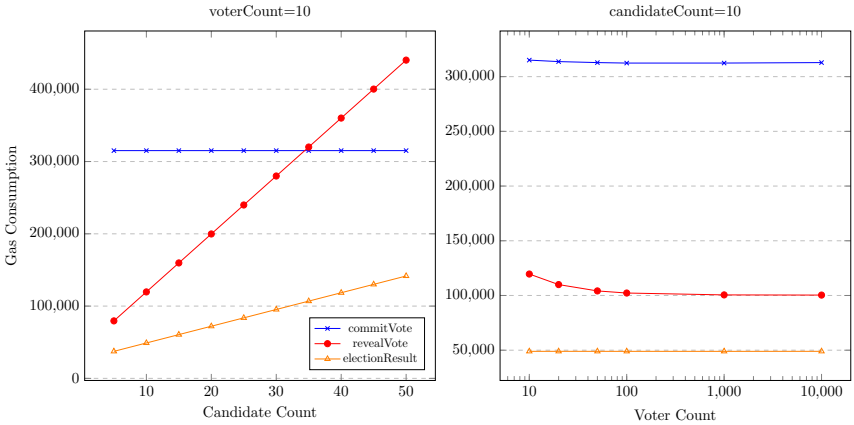}
		\caption{Gas consumption of voter functions}
		\Description[Figure shows two plots for voter function gas consumption]{Plots show gas consumptions of commitVote, revealVote and electionResult functions. The first plot shows gas consumptions of 10 voters with increasing candidate count from 5 to 50. In this plot commitVote stays fixed at 310,000 gas cost, revealVote increases from 90,000 to 450,000 and electionResult increases from 50,000 to 150,000. The second plot shows gas consumptions of 10 candidates with increasing voter count from 10 to 10,000 exponentially. In this plot commitVote stays fixed at 310,000 gas cost, revealVote starts with 150,000 gas then decreases and stays at 100,000 and electionResult is fixed at from 50,000.}
		\label{fig:voter-tests}
			\end{center}
\end{figure}

Gas consumption of the voter functions is given in Figure \ref{fig:voter-tests}. The \textit{commitVote} function has a $O(1)$ gas complexity: it consumes approximately 315.000 gas per transaction. As it verifies the zero-knowledge proofs, it consumes relatively more gas than other functions. The \textit{revealVote} is only affected by the candidate count. The consumption is linear with $O(n_c$) due to the linear time rank-unrank algorithm \cite{permutations}. It can be approximated as $n_c * 8000 + 39000$. 
The voter count does not affect the gas consumption of \textit{revealVote} because each voter acts by himself. However, a down-slope is observed between voterCount=10 and 100. This is due to an additional map initialization cost when \textit{revealVote} is called for the first time. Hardhat gas-reporter takes an average gas consumption if the same function is used more than once. Eventually, the average value is closer to the maximum when there are a few voters and closer to the minimum when there are many voters. The \textit{calculateResult} function is not affected by the voter count, but it is linearly proportional with the candidate count $O(n_c)$. This is due to the \textit{Borda Count} library: the \textit{calculateResult} function iterates over each candidate ID in the tally storage $TS$ to find the candidate with the maximum score. 

\begin{figure}[htbp]
	\begin{center}
		\includegraphics[trim=1cm 17cm 0cm 0cm, clip, width=0.85\textwidth,]{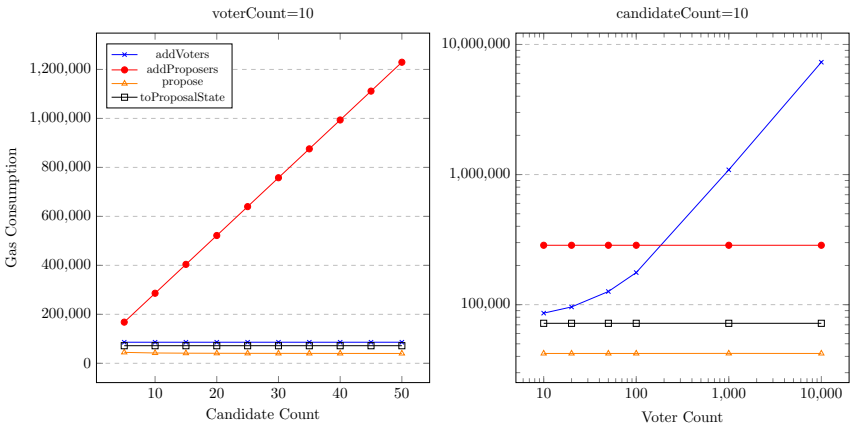}
				\caption{Gas consumption of the EA and Proposer functions}
				\Description[Figure shows two plots for election authority function gas consumption]{Plots show gas consumptions of addVoters, addProposers, propose and toProposalState functions. The first plot shows gas consumptions of 10 voters with increasing candidate count from 5 to 50. In this plot addProposers increases from 200,000 to 1,200,000, and other functions stay fixed at almost 100,000. The second plot shows gas consumptions of 10 candidates with increasing voter count from 10 to 10,000 exponentially. In this plot addVote increases from 100,000 to almost 7,000,000, addProposers is fixed at 250,000, propose is fixed at 50,000 and toProposalState is fixed at 70,000}
				\label{fig:authority-tests}
				\end{center}
\end{figure}

The smart contract deployment costs a total of 3,458,406 gas to EA. Figure \ref{fig:authority-tests} shows gas consumptions of the EA and Proposer functions. Functions for EA are \textit{addVoters}, \textit{addProposers} and \textit{toProposalState}. Proposers use \textit{propose} function. The gas consumption of \textit{addProposers} is linearly proportional to the proposer count. It is approximated as $50180 + (23586 * n_c)$ when all proposers suggest their candidates. The measured gas consumptions of \textit{propose} and the \textit{toProposalState} are $O(1)$ and fairly low. The gas consumption of \textit{addVoters} is directly related to the voter count. 
The block capacity of Ethereum changes between 15 and 20 million gas. 
Experiments show that \textit{addVoters} call exceeds block capacity with approximately 30,000 voters, but EA can issue multiple calls. 
Gas consumption of \textit{addVoters} is measured for the registration of 1000 voters per call in Figure \ref{fig:authority-tests}. 

\begin{figure}[htbp]
	\begin{center}
		\includegraphics[trim=1cm 18.7cm 0cm 0cm, clip, width=0.95\textwidth,]{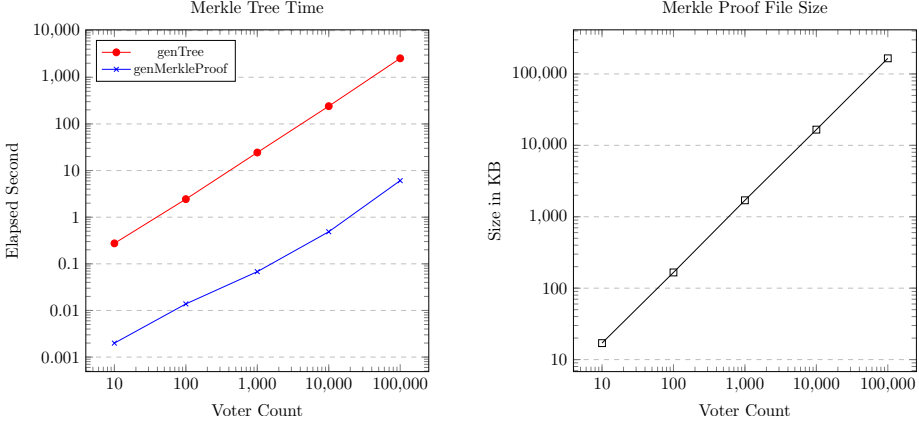}
		\Description[Figure shows two plots for elapsed times and file sizes to generate Merkle trees and proofs]{The first plot shows elapsed times of generating a Merkle tree and generating a Merkle proof with respective to increasing voter counts from 10 to 100,000. The time to generate Merkle tree increases from 0.5 seconds to 42 minutes, generating Merkle proof increases from 1 milliseconds to 10 seconds. The size of the Merkle proof increases from 15 kilobytes to 165 megabytes in respective to increasing voters counts from 10 to 100,000.}
				\caption{Gas consumption of Zero-Knowledge Proofs}
				\label{fig:Merkle-tests}
					\end{center}
\end{figure}

Merkle trees must be generated by both election authority and voters. Figure \ref{fig:Merkle-tests} shows gas consumption for generating Merkle tree (\textit{genTree}) and Merkle proof (\textit{genMerkleProof}). Merkle tree and proof generation have a linear relation with the voter count. The tree can be constructed within 42 minutes for 100,000 voters. Merkle proof generation takes only 6 seconds for 100,000 voters. The Merkle proof file size is 165,5 megabytes for 100,000 voters. The file contains all Merkle proofs for every individual leaf. It means voters can grab the file and find their related Merkle proofs without generating the tree. Table \ref{semaphore-times} presents analyzed the Semaphore circuit setup times, file sizes, and the witness \& proof generation times. Compiling the circuit and generating keys take almost 15 minutes and 250-megabyte file size. This is reasonably feasible as it is a one-time setup only. Each voter uses \textit{genID} and \textit{genIDCommit} functions to generate identity and commitments. In the table, it can be seen that these functions take sub-second times. Voters use \textit{genWitness} and \textit{genProof} to generate their zero-knowledge proofs. Each takes around 10 secs to generate a proof with the Semaphore circuit.

\begin{table}[htbp]
\caption{The Semaphore Circuit \& Function Times}\label{semaphore-times}
\centering
\resizebox{0.6\columnwidth}{!}{
\begin{tabular}[t]{|l|l|l|}
\hline
Operation &
  \begin{tabular}[c]{@{}l@{}}Time \\ (sec)\end{tabular} &
  \begin{tabular}[c]{@{}l@{}}File Size \\ (mb)\end{tabular} \\ \hline
Compile Circuit & 206 & 132 \\ \hline
Key Generation  & 703 & 128 \\ \hline
Generate Verifier Contract &
  0.39 &
  0.01 \\ \hline
\end{tabular}
\begin{tabular}[t]{|l|l|}
\hline
Function              & \begin{tabular}[c]{@{}l@{}}Time \\ (sec)\end{tabular} \\ \hline
genID          & 0.027504                                              \\ \hline
genIDCommit & 0.099805                                              \\ \hline
genWitness            & 1.622                                                 \\ \hline
genProof              & 8.446                                                 \\ \hline
\end{tabular}
}
\end{table}

The Tideman algorithm uses graphs, matrices, and sorting algorithms which are costly for smart contracts. Since vote storage costs are reduced with the aid of the ranking and unranking permutations algorithm, the Tideman library can safely handle 250 voters and 10 candidates. The \textit{calculateResult} function hits the 30 million gas limit with more than 250 voters. This makes it a viable option for small-scale elections. The gas consumptions of both Borda Count and Tideman libraries  for 250 voters and 10 candidates are presented in Table \ref{tideman-results}. The \textit{electionResult} and \textit{revealVote} functions discloses the difference between to tallying algorithms in terms of gas costs. Other differences are due to the in contract sizes of used libraries. Tideman library is bigger in contract size than the Borda count library.   
\begin{table}[h]
\caption{Gas Consumption of Borda Count and Tideman Libraries (voterCount=250, candidateCount=10)}
\label{tideman-results}
\begin{center}
\resizebox{0.4\linewidth}{!}{
\begin{tabular}{|l|r|r|}
\hline
Transaction     &Borda & Tideman \\ \hline
Deployment      &3,523,900 & 4,135,997\\ \hline
addVoters       &325,142 & 325,166 \\ \hline
addProposers    &286,040 & 286,064 \\ \hline
propose         &42,681 & 42,693    \\ \hline
toProposalState &71,877 & 71,877    \\ \hline
commitVote      &312,405   & 312,404   \\ \hline
revealVote      &103,334 & 88,844    \\ \hline
electionResult  & 48,953 & 22,210,898\\ \hline
\end{tabular}
}
\end{center}
\end{table}

\subsection{Comparison with Existing Work}
ElectAnon is compared with McCorry et al.\cite{ovnet} and PriScore \cite{priscore}. McCorry et. al. \cite{ovnet} supports yes/no election. Hence, they conducted their gas consumption tests with 40 voters. PriScore \cite{priscore} shows gas costs for each of their different functions. Its total gas consumption is calculated for 40 voters and 10 candidates. Since the gas consumption of {\it Setup} state is not mentioned, it is skipped. Then, ElectAnon is executed for an election with 40 voters and 10 candidates for a fair comparison. The results are presented in Table \ref{gas-comparison}. It shows that ElectAnon offers an 83\% and 89\% decrease in total election gas consumption compared to McCorry et al. \cite{ovnet} and Priscore \cite{priscore}, respectively. 

Further analysis can be done with respect to Ethereum's block capacity and size. At the time of McCorry et. al. writing their paper \cite{ovnet}, they hit an Ethereum block capacity (2 million gas) for a single vote in a 50-voter setup. Currently, the block capacity is 30 million gas on average, which makes the work support around 650 voters at maximum as of today. The cost of casting a single vote is shown as approximately 3,300,000 gas which would be equivalent to almost \$900 today, which is not feasible. Panja et. al. \cite{panja} extends McCorry et. al.'s work, but it turns more expensive: even a single voting cost exceeds the block limit of 8 million. They tried to split the voting transaction into five sub-transactions, but in the end, a single voting transaction costs 40,102,222 gas. In \cite{YANG2020859}, one vote size is measured as 100 KB in a setup with 15 candidates. This does not fit into one Ethereum block, which is almost 80 KB on average at the time of writing their paper. In Priscore \cite{priscore}, gas consumption in Commit and Vote states is dependent on the number of voters and candidates. When the total gas consumption is computed for ten candidates and 50 voters, Commit and Vote states consume a total of almost 1,100,000 and 3,500,000 gas, respectively.  In ElectAnon, gas consumption of Commit and Reveal functions are independent of the number of voters and candidates. A total of nearly 420,000 gas is consumed per vote. This shows that ElectAnon is ten times more efficient than its closest rival, Priscore, in a 50-voter 10-candidate setup.
\begin{table}[htbp]
\caption{Gas Consumption Comparison (voterCount=40, candidateCount=10)}\label{gas-comparison}
\centering
\resizebox{0.6\columnwidth}{!}{
\begin{tabular}{|l|r|r|r|}
\hline
Entity: Transaction & McCorry et al.\cite{ovnet} & Priscore \cite{priscore}   & This Work   \\ \specialrule{.1em}{.05em}{.05em} 
A: Deploy           & 6,215,811      & -           & 3,430,754  \\ \hline
A: Add Voters       & 2,153,461      & -           & 113,963    \\ \hline
A: Add Proposers    & -              & -           & 286,040    \\ \hline
A: State Change     & 3,320,433      & -           & 71,877     \\ \hline
P: Propose          & -              & -           & 42,681     \\ \hline
V: Register         & 763,118        & -           & -          \\ \hline
V: Commit           & 70,112         & 1,107,374   & 312,856    \\ \hline
V: Vote             & 2,490,412      & 3,579,468   & 105,140    \\ \hline
E: Tally            & 746,485        & 60,096      & 48,937     \\ \hline
\rowcolor[gray]{.9} 
Authority Total     & 12,436,190     & 60,096           & 3,665,531  \\ \hline
\rowcolor[gray]{.9} 
Proposer Total      & -              & -           & 42,681     \\ \hline
\rowcolor[gray]{.9}
Voter Total         & 3,323,642      & 4,686,842   & 417,996    \\ \hline
\rowcolor[gray]{.8}
Election Total      & 145,381,870    & 187,533,776 & 20,812,181 \\ \hline
\end{tabular}
}
\\{\tiny A: Authority, P: Proposer, V: Voter, E: Everyone}
\end{table}

\subsection{Scalability Analysis}
Gas consumption is a very strong indicator of scalability since there is a limit on the maximum consumable gas in Ethereum. This limit dynamically changes with the network state and Ethereum fork version. Measuring the transaction delays and finalization times is not very reasonable as they completely depend on the network state and the underlying blockchain platform. It could be misleading to measure the scalability of a smart contract, or generally a decentralized application, with transaction and finalization times.

At the time of writing this work, the gas price in Ethereum is equivalent to nearly 100 \textit{gwei}, and one ETH price is approximately \$4500. A gwei equals $10^9$ ETH. Hence, a single gas costs $100 * \$0.0000045 = \$0.00045$. In this work, total gas consumption of a voter is 417,996 which makes $\$0.00045  * 417,996 = \$188.06$. ElectAnon can execute on every Ethereum Virtual-Machine compatible network such as \textit{Avalanche}. Avalanche implements a novel consensus mechanism with proof-of-stake Sybil protection \cite{rocket2018snowflake}. It offers a faster finalization time with fairly low gas fees compared to Ethereum. The Avalanche gas price changes between 25-150 \textit{nAVAX} (equivalent to \textit{gwei}) \cite{noauthor-transaction-nodate}. Average current price of Avalanche is \$85. If gas price is assumed to be 100 nAVAX, voting transactions costs on Avalanche can be as low as $100 / 10^9 * \$85 * 417,996 = \$3.55$. An election with 10 voters and 10 candidates is tested in Avalanche local network. It has demonstrated that ElectAnon is compatible with the Avalanche network with the same gas consumption as the Ethereum network. However, since gas prices and the AVAX price are much lower than the Ethereum main network, it significantly reduces the election cost.  

The experiments have shown that voter functions, i.e. \textit{commitVote}, \textit{revealVote}, \textit{calculateResult}, are not affected by the increased voter count. It means that voters do not pay for extra gas in a large-scale election. The cost of election authority increases with both candidate count for \textit{addProposers} and voter count for \textit{addVoters}. Hence, the election authority should have enough resources to start an election. A possible bottleneck of ElectAnon could be adding voters to the eligible list. It costs almost a total of 10,000,000 gas for 10,000 voter registration, which costs almost $\$5000$ in Ethereum. It means that a single leaf insertion costs approximately $5000/10,000 = \$0.5$. This is a fair cost for small to medium-scale elections, i.e. up to 10,000 voters. The gas cost increases linearly. A large-scale election with 1,000,000 voters would cost approximately \$500.000. Even though ElectAnon offers the best gas consumption among the existing studies, \$500.000 is still too much. The total election cost can be decreased to $\$8500$ in the \textit{Avalanche} network due to its significantly lower gas price. 

\subsection{Merkle Forest} \label{Merkle-forest-exp}

Tests for the Merkle forest extension are carried out with various fixed tree sizes. Each size represents the total leaf size, and thus the maximum voter count can fit into one tree. Results for two different implementations are shown in Table \ref{table:Merkle-forest}. One of the implementations uses \textit{the Keccak-256} hash function; the other one uses public input lists. Implementation with the Keccak-256 hash function consumes less gas as the smart contract takes a single hash input. In the implementation without Keccak-256, all leaves are passed to the verifier to be verified. Giving a complete list instead of a single hash increases gas consumption with respect to the increased voter count. The implementation without the Keccak-256 hash function produces fewer circuit constraints. As a result, generation times and file sizes are much lower. 

The circuit with \textit{Keccak-256} is more feasible for a smart contract as it consumes almost one-quarter of gas compared to the one without Keccak-256. Inserting a 256-sized tree with the Keccak-256 circuit consumes 482,409 gas, whereas the implementation without Merkle forest consumes 276,929 for 200 voter registration. Hence, the Merkle forest extension can be a feasible option for smaller-scale elections.  It's possible to generate a 256-size tree circuit and insert multiple trees into the smart contract to increase the total voter count in successive rounds. 

\begin{table}[htbp]
\caption{Test Results for Merkle Forest Implementations}
\label{table:Merkle-forest}
\centering
\resizebox{\columnwidth}{!}{
\begin{tabular}{@{}llllllllllll@{}}
\toprule
  \begin{tabular}[c]{@{}l@{}}Used \\ Keccak\end{tabular} & 
  Size &
  Constraints &
  \begin{tabular}[c]{@{}l@{}}Compile \\ Time\end{tabular} &
  \begin{tabular}[c]{@{}l@{}}Setup \\ Time\end{tabular} &
  \begin{tabular}[c]{@{}l@{}}Witness \\ Time\end{tabular} &
  \begin{tabular}[c]{@{}l@{}}Proof \\ Time\end{tabular} &
  \begin{tabular}[c]{@{}l@{}}Deploy \\ Gas\end{tabular} &
  \begin{tabular}[c]{@{}l@{}}Insert \\ Gas\end{tabular} &
  \begin{tabular}[c]{@{}l@{}}Insert Gas per \\ Size\end{tabular} &
  \begin{tabular}[c]{@{}l@{}}Compiled \\ Size \\ (MB)\end{tabular} &
  \begin{tabular}[c]{@{}l@{}}Proving  \\ Key Size \\ (MB)\end{tabular} \\ \midrule
Yes & 16  & 661,908   & 0:00:29 & 0:04:44 & 0:00:10 & 0:00:30 & 1,404,004 & 322,106  & 20,131.625   & 452  & 229  \\
No  & 16  & 39,601    & 0:00:12 & 0:00:24 & 0:00:07 & 0:00:11 & 1,714,373 & 419,031  & 26,189.4375  & 292  & 14   \\ \midrule
Yes & 32  & 1,326,724  & 0:01:05 & 0:09:42 & 0:00:22 & 0:01:07 & 1,405,936 & 332,534  & 10,391.6875  & 957  & 459  \\
No  & 32  & 81,874    & 0:00:27 & 0:00:53 & 0:00:16 & 0:00:25 & 2,128,664 & 558,448  & 17,451.5     & 636  & 28   \\ \midrule
Yes & 64  & 2,656,356  & 0:02:53 & 0:21:06 & 0:00:53 & 0:02:28 & 1,404,004 & 353,931  & 5530.171875 & 1900 & 918  \\
No  & 64  & 166,321   & 0:01:02 & 0:01:52 & 0:00:34 & 0:00:53 & 2,948,257 & 839,127  & 13,111.35938 & 1300 & 57   \\ \midrule
Yes & 128 & 5,162,019  & 0:05:18 & 0:37:37 & 0:01:36 & 0:04:29 & 1,404,016 & 396,685  & 3099.101563 & 3900 & 1800 \\
No  & 128 & 335,281   & 0:02:16 & 0:03:17 & 0:01:08 & 0:01:48 & 4,595,134 & 1,405,349 & 10,979.28906 & 2700 & 114  \\ \midrule
Yes & 256 & 10,173,347 & 0:30:49 & 1:16:15 & 0:03:17 & 0:10:03 & 1,404,640 & 482,409  & 1884.410156 & 7800 & 3500 \\
No  & 256 & 673,201   & 0:07:56 & 0:06:48 & 0:02:17 & 0:03:55 & 7,890,686 & 2,559,015 & 9996.152344 & 5400 & 228  \\ \bottomrule
\end{tabular}
}
\end{table}

\section{Conclusion}

In this work, ElectAnon, a blockchain-based, anonymous ranked-choice voting protocol is proposed. It ensures anonymity with zero-knowledge proofs. A zero-knowledge gadget, \textit{Semaphore} \cite{semaphore-paper} is utilized as it provides anonymous membership proofs. It is designed to be robust and uninterruptible in the voting phase. The protocol not only assures critical election requirements but also scales to be used in large-scale elections. The protocol makes use of a linear-time algorithm \cite{permutations} for encoding and decoding ranked-choice ballots. We also defined some of the most critical election requirements while providing a deep-down analysis of prior works and ElectAnon. 

ElectAnon is implemented with Ethereum smart contracts to ensure decentralization and robustness. Gas costs for different implementation options are analyzed with experiments and compared with related work. The financial analysis of the proposal also shows that ElectAnon can especially be beneficial for governance applications like decentralized autonomous organizations (DAOs).



\begin{acks}
This research was partially supported by Bo\u{g}azi\c{c}i University Scientific Research Projects (BAP) No: 17A01P7.
\end{acks}

\bibliographystyle{ACM-Reference-Format}
\bibliography{base}

\appendix

\section{Tallying Algorithms}
\label{chapter:tallying-algorithms}

\SetKwComment{Comment}{// }{}
\SetKwInOut{KwIn}{Inputs}
\SetKwInOut{KwOut}{Output}
\SetKwInOut{Ensure}{Ensure}
\SetKwInOut{Require}{Require}
\SetKwInOut{Storage}{Storage}
\SetKw{KwBy}{by}
\RestyleAlgo{ruled}

\begin{figure}[htbp]
\begin{minipage}[t]{0.5\linewidth}
  \vspace{0pt}  
  \begin{algorithm}[H]
\caption{Borda Count Tally}\label{alg:two}
\KwIn{$v$: (uint) the vote ID (rank) \\
$P$: (uint) count of proposals (candidates)}
\Storage{$\textit{VC}$: (map\{uint:uint\}) stored vote counts}
$\pi \gets [0, 1, \cdot\cdot\cdot, n]$ \Comment*[r]{identity vector}
$vec \gets unrank(v, P, \pi)$\;
\For{$i\gets0$ \KwTo $vec.length$ \KwBy $1$}{
$VC[vec[i]] \mathrel{+}= v.length - i$\;
}
\end{algorithm}
\end{minipage}%
\begin{minipage}[t]{0.5\linewidth}
  \vspace{0pt}  
  \begin{algorithm}[H]
\caption{Borda Calculate Result}\label{alg:three}
\KwIn{$P$: (uint) count of proposals (candidates)
}
\KwOut{$w$: the winner candidate ID}
\Storage{$\textit{VC}$: (map\{uint:uint\}) stored vote counts}
$max \gets 0$\;
$w \gets 0$\;
\For{$i\gets0$ \KwTo $P$ \KwBy $1$}{
\If{$VC[i] > max$}
{
 $max \gets VC[i]$\;
 $w \gets i$\;
}
\textit{return w}}
\end{algorithm}
\end{minipage}%
\end{figure}

\begin{figure}[htbp]
\begin{minipage}[t]{0.5\linewidth}
  \vspace{0pt}  
  \begin{algorithm}[H]
\caption{Tideman Tally}\label{alg:four}
\KwIn{$v$: (uint) the vote ID (rank) \\ 
$P$: (uint) count of proposals (candidates)}
\Storage{$\textit{VC}$: (map\{uint:uint\}) stored vote counts\\
$\textit{RL}$: (uint[]) unique rank IDs}
\If{VC[v] == 0} {
    $RL\textit{.push}(v)$
}
$VC[v] \gets VC[v]+1$\;
\end{algorithm}
\end{minipage}%
\begin{minipage}[t]{0.5\linewidth}
  \vspace{0pt}  
  \begin{algorithm}[H]
\caption{Tideman Calculate Result}\label{alg:five}
\KwIn{$P$: (uint) count of proposals (candidates)\\
$\textit{VC}$: (map\{uint:uint\}) stored vote counts \\
$\textit{RL}$: (uint[]) unique rank IDs
}
\KwOut{$w$: (uint) the winner candidate ID}
$\textit{prefs} \gets \textit{getPreferenceMatrix(P, VC, RL)}$\;
$\textit{locked} \gets \textit{getLockedPairs(P, prefs)}$\;
\For{$i\gets0$ \KwTo $P$ \KwBy $1$}{
$\textit{source} \gets true$\;
\For{$j\gets0$ \KwTo $P$ \KwBy $1$}{
\If{\textit{locked}[j+1][i+1]}{
$\textit{source} \gets false$\;
break\;
}
}
\If{source == true} {
                \textit{return i+1};
            }
            }
\textit{return 0} \Comment{Winner could not be determined}

\end{algorithm}
\end{minipage}
\end{figure}

\begin{figure}[htbp]
{
\setlength{\interspacetitleruled}{0pt}%
\setlength{\algotitleheightrule}{0pt}%
\begin{algorithm}[H]

    \SetKwFunction{FMain}{getPreferenceMatrix}
    \SetKwProg{Fn}{Function}{:}{}
\Fn{\FMain{$P$, $VC$, $RL$}}{
\For{$j\gets0$ \KwTo $RL.length$ \KwBy $1$}{
$counter \gets 0$\;
$rank \gets RL[j]$\;
$voteCount \gets VC[rank]$\;
$vec \gets unrank(rank, P)$\;
\For{$i\gets0$ \KwTo $P$ \KwBy $1$}{
\For{$k\gets0$ \KwTo $i+1$ \KwBy $1$}{
$\textit{prefs}[counter][0] \gets i +1$\;
$\textit{prefs}[counter][1] \gets k +1$\;
\uIf{$v.\textit{indexOf}(i+1) < v.\textit{indexOf}(k+1)$}{
$\textit{prefs}[counter][2] \gets voteCount$\;
} \Else{
$\textit{prefs}[counter][3] \gets voteCount$\;
}
$counter \gets counter + 1$\;
}
}
}
\textit{return prefs}\;
}
\end{algorithm}

\begin{minipage}[t]{0.5\linewidth}
  \vspace{0pt}  
  \begin{algorithm}[H]
\SetKwFunction{FMain}{getLockedPairs}
\SetKwProg{Fn}{Function}{:}{}
\Fn{\FMain{$P$, $\textit{prefs}$}}{
$\textit{pairs} \gets \textit{getPairs(P, prefs)}$\;
$\textit{sortedPairs} \gets \textit{insertionSort(pairs)}$\;
\For{$i\gets0$ \KwTo $sortedPairs.length$ \KwBy $1$}{
$\textit{winner} = sortedPairs[i][0]$\;
$\textit{loser} = sortedPairs[i][1]$\;
\If{$\textit{hasNoCycle(winner,
                    loser,
                    P,
                    locked)}$}
                    {
$\textit{locked[winner][loser]} \gets true$
}
}
\textit{return locked}\;
}
\end{algorithm}

\end{minipage}%
\begin{minipage}[t]{0.5\linewidth}
  \vspace{0pt}
\begin{algorithm}[H]

\SetKwFunction{FMain}{getPairs}
\SetKwProg{Fn}{Function}{:}{}
\Fn{\FMain{$P$, $\textit{prefs}$}}{
\For{$j\gets0$ \KwTo \textit{prefs.length} \KwBy $1$}{
\uIf{$\textit{prefs}[i][2] > \textit{prefs}[i][3]$}{
$\textit{pairs[i][0]} \gets \textit{prefs[i][0]}$\;
$\textit{pairs[i][1]} \gets \textit{prefs}[i][1]$\;
$\textit{pairs[i][2]} \gets \textit{prefs}[i][2] - \textit{prefs}[i][3]$\;
}
\ElseIf{$\textit{prefs}[i][2] < \textit{prefs}[i][3]$}{
$\textit{pairs[i][0]} \gets \textit{prefs[i][1]}$\;
$\textit{pairs[i][1]} \gets \textit{prefs}[i][0]$\;
$\textit{pairs[i][2]} \gets \textit{prefs}[i][3] - \textit{prefs}[i][2]$\;
}
}
\textit{return pairs}\;
}
\end{algorithm}
\end{minipage}
}
\end{figure}

\end{document}